\documentclass[prl,twocolumn,superscriptaddress]{revtex4-1}
\usepackage{amssymb}
\usepackage{amsfonts}
\usepackage{bbm}
\usepackage{mathrsfs}
\usepackage{graphics,graphicx,epsfig,bm,amsmath,amsthm,amssymb}
\usepackage{bm}
\usepackage{bbm}
\usepackage{longtable}
\usepackage{multirow}
\usepackage{array}
\usepackage{color}
\usepackage{cases}
\usepackage{booktabs}
\usepackage[usenames,dvipsnames]{xcolor}

\usepackage{xcolor}
\usepackage{float}
\usepackage[a4paper,colorlinks=true,
linkcolor=blue,citecolor=blue,
pdfauthor={ },
pdftitle={ },
pdfsubject={ },
pdfkeywords={ }]{hyperref}

\newcommand{\Rmnum}[1]{\expandafter\@slowromancap\romannumeral #1@}

\begin{document}
	\title{$T_2$-limited dc Quantum Magnetometry via Flux Modulation}
	
	%\baselineskip24pt
	
	% Make the title.
	
	%\maketitle
	
	\affiliation{CAS Key Laboratory of Microscale Magnetic Resonance and School of Physical Sciences, University of Science and Technology of China, Hefei 230026, China}
	\affiliation{Hefei National Laboratory, Hefei 230088, China}
	\affiliation{CAS Center for Excellence in Quantum Information and Quantum Physics, University of Science and Technology of China, Hefei 230026, China}
	%\author{Yijin Xie$^{1,3}$, Caijin Xie$^{1,3}$, Yunbin Zhu$^{1,3}$, Ke Jing$^{1,3}$, \\
	%            Yu Tong$^{1,3}$, Xi Qin$^{1,3}$, Haosen Guan}
	\author{Yijin Xie}
	\author{Caijin Xie}
	\author{Yunbin Zhu}
	\author{Ke Jing}
	\author{Yu Tong}
	\affiliation{CAS Key Laboratory of Microscale Magnetic Resonance and School of Physical Sciences, University of Science and Technology of China, Hefei 230026, China}
	\affiliation{CAS Center for Excellence in Quantum Information and Quantum Physics, University of Science and Technology of China, Hefei 230026, China}
	
	\author{Xi Qin}
	\affiliation{CAS Key Laboratory of Microscale Magnetic Resonance and School of Physical Sciences, University of Science and Technology of China, Hefei 230026, China}
	\affiliation{Hefei National Laboratory, Hefei 230088, China}
	\affiliation{CAS Center for Excellence in Quantum Information and Quantum Physics, University of Science and Technology of China, Hefei 230026, China}
	
	\author{Haosen Guan}
	\affiliation{CAS Key Laboratory of Microscale Magnetic Resonance and School of Physical Sciences, University of Science and Technology of China, Hefei 230026, China}
	\affiliation{CAS Center for Excellence in Quantum Information and Quantum Physics, University of Science and Technology of China, Hefei 230026, China}
	\author{Chang-Kui Duan}
	\author{Ya Wang}
	\author{Xing Rong}
	\email{xrong@ustc.edu.cn}
	\author{Jiangfeng Du}
	\email{djf@ustc.edu.cn}
	\affiliation{CAS Key Laboratory of Microscale Magnetic Resonance and School of Physical Sciences, University of Science and Technology of China, Hefei 230026, China}
	\affiliation{Hefei National Laboratory, Hefei 230088, China}
	\affiliation{CAS Center for Excellence in Quantum Information and Quantum Physics, University of Science and Technology of China, Hefei 230026, China}

	\begin{abstract}
		%Due to multiple noise sources in the real physical systems,
		%This limits the potential applications in the field of magnetic field detection.
		%A proposed method to enhanced the sensitivity is to transfer the low-frequency magnetic field to high-frequency, called flux modulation.
		%sensitivity of most magnetometers, including the ,%However, there is little work to demonstrate the advantages of this method in the field of quantum sensing.
		
		High-sensitivity magnetometry is of critical importance to the fields of biomagnetism and geomagnetism.
		However, the magnetometry for the low-frequency signal detection meets the challenge of sensitivity improvement, due to multiple types of low-frequency noise sources.
		In particular, for the solid-state spin quantum magnetometry, the sensitivity of low frequency magnetic field has been limited by short $T_2^*$.
		Here, we demonstrate a $T_2$-limited  dc quantum magnetometry based on the nitrogen-vacancy centers in diamond.
		The magnetometry, combining the flux modulation and the spin-echo protocol, promotes the sensitivity from being limited by $T_2^*$ to $T_2$ of orders of magnitude longer.
		The sensitivity of the dc magnetometry of 32 $\rm pT/Hz^{1/2}$ has been achieved, overwhelmingly improved by 100 folds over the Ramsey-type method result of 4.6 $\rm nT/Hz^{1/2}$. Further enhancement of the sensitivity have been systematically analyzed, although challenging but plenty of room is achievable. Our result sheds light on realization of room temperature dc quantum magnetomerty with femtotesla-sensitivity in the future.
		%, which is based on a solid-state spins and mechanical resonator hybrid system.
		%In this work, the low-frequency magnetic field is transferred to the 10.795 kHz through the mechanical resonator. The limitation of $T_2^*$ for low-frequency magnetometry is broken and advanced to $T_2$.
		%In the future, the method can be further optimized and achieves a low-frequency sensitivity of femtotesla.
		% paves a way for the in the future,has taken
	\end{abstract}
	
	\maketitle
	
	Sensing weak magnetic fields plays an important role in many  frontier science and technology, such as biological signal measurement \cite{Boto2018,Cohen1967}, geomagnetic survey \cite{Lenz2006}, magnetic imaging applications \cite{Glenn2017}, and exotic interactions examination \cite{Jiao2021}.
	Quantum sensors have been proposed with high sensitivity and other advantages \cite{Degen2017,Budker2007,Dolabdjian2017,Dolabdjian2017}. In particular, a type of solid-state spin system,
	the electron spins of Nitrogen-Vacancy (NV) centers in diamond, has been proposed to be utilized as a quantum magnetometry with potential sensitivity up to femtotesla at room temperature \cite{Taylor2008}.  Since then, great efforts have been paid to enhance its sensitivity for detecting static or low-frequency magnetic fields, including cavity-enhanced infrared absorption magnetometry \cite{Jensen2014, Chatzidrosos2017}, the total internal reflection \cite{Clevenson2015}, flux concentration \cite{Xie2021,Fescenko2020}, and continuously excited Ramsey measurements combined with lock-in detection \cite{Zhang2021c}.
	The NV-based magnetometry has been improved to be sub-picotesla sensitivity \cite{Xie2021,Fescenko2020} with flux concentration and picotesla sensitivity\cite{Barry2017} without flux concentration.
	
	There is still a great gap between the current low-frequency sensitivity and  the expected femotesla-sensitivity for NV-based magnetometry. The main obstacle is the short $T_2^*$ of the ensemble NV center system, which essentially limits sensitivities of the continuous wave type and Ramsey-type NV-based magnetometers.
	Recently, intense efforts have been paid to increase $T_2^*$ by improving the material of the diamond or by developing advanced quantum control technologies \cite{Barry2020,Bauch2018,Wolf2015, Bauch2018, Zhang2021c}.
	Schemes have been proposed to break the $T_2^*$ limit, including the ancilla-assisted frequency up-conversion \cite{Liu2019a} and fast rotation of diamonds \cite{Wood2018a}.
	By dynamically engineering quantum states of NV centers, these works turn the static magnetic field into a pseudo ac one. However, significant sensitivity advancement still remains elusive.

	In this work, a dc quantum magnetometry with $T_2$-limited based on the ensemble of NV centers in diamond is demonstrated.
	A Flux Concentration and Modulation (FCM) technique has been utilized to transfer the dc magnetic signal to an ac one.
	Then the ac signal can be detected with the spin-echo sequence \cite{Hahn1950} applied on the NV magnetometry.
	Our work has achieved a $\sim$140 folds sensitivity enhancement from 4.6 $\rm nT/Hz^{1/2}$ to 32 $\rm pT/Hz^{1/2}$.
	The low-frequency sensitivity limit has been successfully boosted from $T_2^*$ to $T_2$.
	%It is a big challenge for the sensitivity enhancement compared to Ramsey protocol before, since the previous scheme.
	%With the state-of-art technologies, the sensitivity of the magnetometry can be further extended to the level of femtotesla.
	We also systematically analyzed limitations for enhancement of the sensitivity with the current magnetometry, and our result paves a way towards future realization of room temperature dc quantum magnetomerty with femotesla-sensitivity.
	%Spin bath noise caused by impurities like $^{13}$C, $N_{s}$ and other paramagnetic centers will lead to decoherence of NV electronic spin.
	\begin{figure*}
		\includegraphics[width=1.0\textwidth]{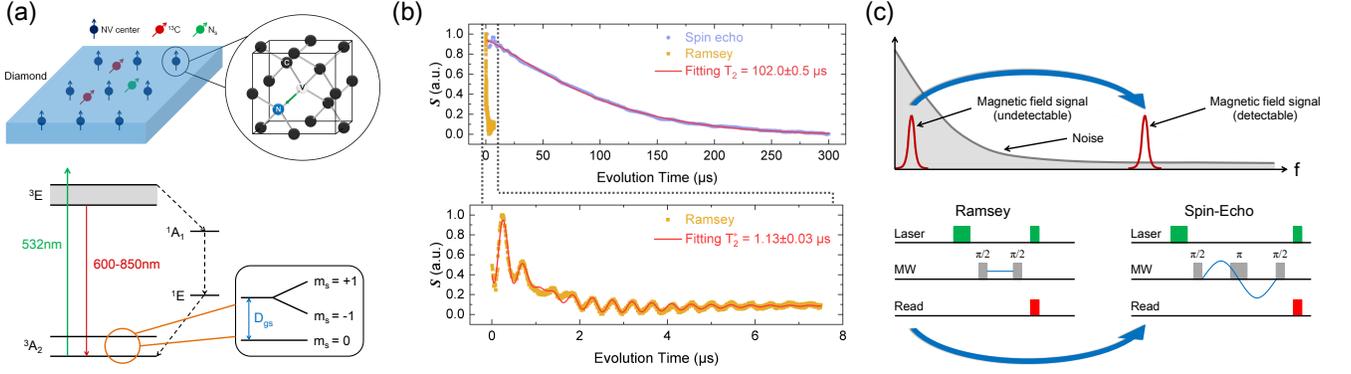}% Here is how to import EPS art
		\caption{\label{fig:1}The system of  the NV centers in diamond and the basic idea of our scheme. (a) Diagram of diamond containing NV centers and other impurities. $^{13}$C and N$_{s}$ represent for isotope of carbon atom and isolated substitutional nitrogen atom in diamond lattice, respectively. (b) Experimental results of Ramsey experiment and spin-echo experiment. The data of Ramsey experiment is fitted by exponential damping and sum of three cosine function with different angular frequency. $T_{2}^*=1.13\pm0.03~\rm \mu s$ is obtained. The oscillation in $T_{2}^*$ data is induced by the hyperfine interaction between NV and $^{14}$N. The data of spin-echo signal is fitted by exponential damping function and get $T_{2}=102.0\pm0.5~\rm \mu s$. (c) The schematic of the magnetic field modulation. The raw external signal is in the low-frequency domain originally, while the low-frequency noise and spin bath noise are dominant. The modulation method transfers the raw signal into high-frequency domain. The noise in the high-frequency domain presents less intensity compared to the low-frequency domain. }
	\end{figure*}
	
	%come from the $\pi$ pulse in dynamic decoupling sequence ;;;difference of $T_2^*$ and $T_2$ is about 1-2 orders of magnitude
	%, and being further used in the AC magnetic field measurement
	%So the ground state is splited into one m_s=0 state and  m_s=\pm 1 states which are D_gs=2.87GHz higher in frequency.
	%the vibronic states by above-band excitation, after which  is brought quickly into electronic excited states by phonon relaxation
	The NV center consists of a substitutional nitrogen adjacent to a lattice vacancy.
	The energy levels of NV center is shown in Fig. \ref{fig:1}(a).
	The ground state of NV center is a spin triplet state $^3A_2$, which contains sublevels of $m_s=0$ and $m_s=\pm 1$ \cite{Barry2020}.
	The excited state of NV center includes some discrete electronic states and continuous vibronic states with higher energy.
	A 532 nm laser can be used to excite the NV center.
	Then the NV center decays to the ground state with fluorescence emitting from 600 to 850 nm \cite{Barry2020}.
	The intensity of fluorescence is dependent on the spin state of NV center in ground state.
	Thus the readout of NV center can be realized by the 532 nm laser excitation and fluorescence acquisition.
	
	%The excited state can also transit to m_s=\pm 1 states through metastable spin singlet state 1A_1 and no phono is emitted in this situation.
	
	There are many types of impurities in the diamond, such as $^{13}$C and N$_{s}$ shown in Fig. \ref{fig:1}(a).
	The inevitable interactions between these impurities and the electron spin of NV center are the main contributions to the short dephasing time $T_2^*$. %\cite{Bauch2018} .
	%The impurities and the imhomogeneitieslead to the decoherence of the NV centers.
	In our experiment, the $T_2^*$ is $1.13\pm0.03~\rm \mu s$ (Fig. \ref{fig:1}(b)).
	This dephasing effect can be greatly suppressed by dynamical decoupling technique such as spin-echo sequence, and the coherence time of the NV centers can be prolonged to $T_2$.
	The prolonged coherence time $T_2$ with spin-echo sequence is $102.0\pm0.5~\rm \mu s$ (Fig. \ref{fig:1}(b)).
	Unfortunately, the dynamical decoupling technique also cancels the effect of dc magnetic field on NV centers. Therefore,  the prolonged coherence time via dynamical decoupling technique cannot ensure the improvement of the dc magnetic sensitivity of the NV magnetometry.
	%So the coherence time can be prolonged by the dynamic decoupling sequence \cite{Du2009}.
	%In this work, we measure the coherence time and obtain $T_2^*$ of $1.13\pm0.03~\rm \mu s$ and $T_2$ of $102.0\pm0.5~\rm \mu s$ shown in Fig. \ref{fig:1}(b).
	In this work, we utilized the FCM method to transfer the low-frequency signal into the high-frequency domain, while the Ramsey-type protocol can be replaced by the spin-echo protocol accordingly, as demonstrated in Fig. \ref{fig:1}(c).
	Our method will result in an extending limit to $T_2$ for the low-frequency magnetic field detection.
	% \cite{Barry2017,Chatzidrosos2017,Zhang2021c}.
	%the dynamic decoupling sequence cannot be used in the detection of the DC magnetic field, since .
	%$T_2$ is about 2 orders of magnitude longer than $T_2^*$.
	%However, the $\pi$ pulse in dynamic decoupling sequence also leads to the cancellation of the low-frequency signal during phase accumulation.
	
	\begin{figure*}
		\includegraphics[width=1.0\textwidth]{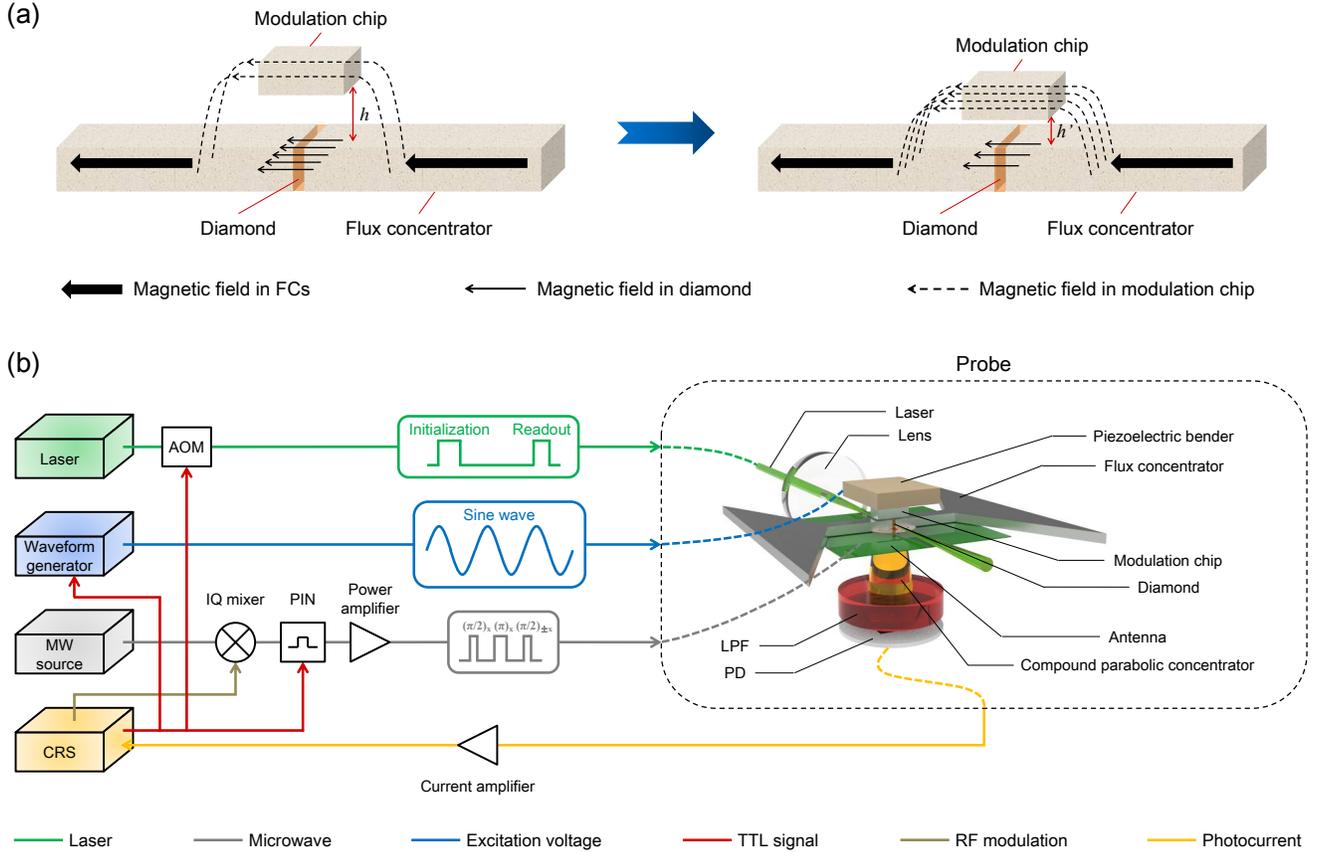}% Here is how to import EPS art
		\caption{\label{fig:2}Schematic of experimental setup. (a) Principle of FCM technique. Bold arrow, light arrow and dashed arrow represent for the magnetic field in FCs, in interval between FCs and in modulation chip, respectively. Thickness of arrow indicates the intensity of magnetic field. $h$ represents for the gap between modulation chip and FCs.  With $h$ decreasing to $h'$, the intensity of magnetic field in modulation chip increases, and the intensity of magnetic field in interval between FCs decreases conversely. (b) Experimental setups. CRS represents for home-built control and readout system. The CRS controls output of laser and microwave. The experiment is conducted by spin-echo sequence with $\pi/2-\pi-\pi/2$ microwave pulses. Waveform generator provides sinusoid signal for the excitation of piezoelectric bender. The output of the waveform generator is triggered by CRS. Dashed box on the right side is 3D rendering of the probe. LPF is low-pass filter. PD is photo diode. Fluorescence emitted from diamond is collected by a compound parabolic concentrator and finally detected by a PD. The photocurrent is recorded by CRS for analysis. Meanings of lines with different colors are shown below.}
	\end{figure*}

	% the motion of the film leads to the variation of the magnetic field near the film.
	%To break $T_2^*$'s limit in low-frequency signal detection, , and the decrease in the diamond
	%The FCM method is used to transfer the magnetic field into the high-frequency domain in our work.
	We utilized the mechanical motion scheme \cite{Tian2013, Edelstein2006,Guedes2008} to realize the FCM, as shown in Fig. \ref{fig:2}(a).
	The mechanical motion scheme is implemented by a piezoelectric bender integrated with a modulation chip  made of high permeability permalloy.
	The diamond sample is clamped by two Flux Concentrators (FCs).
	The modulation chip is placed on the top of the diamond sample and FCs.
	The FCs result in magnification of magnetic field on the diamond.
	The reduction of the gap $h$ leads to the decrease of magnetic field intensity in the diamond \cite{Tian2013}.
	Thus, with the periodically driven modulation chip, the magnetic field intensity in the diamond will become time-dependent.
	The external dc magnetic field can be modulated to an ac magnetic field, which has the same frequency as the driven signal, and it can be then detected by the NV magnetometry with spin-echo pulse sequence.
	This scheme provides a potential improvement in sensitivity by a factor of $\sqrt{T_{2} / T_{2}^*}$.
	
	In our experiment, the resonant frequency of the piezoelectric bender is about 10.8 kHz with the load of modulation chip (See Sec. I in Supplementary Material).
	The maximum vibration amplitude of the piezoelectric bender is 3.6 $\rm \mu m$.
	%The moving distance of the modulation chip is up to nearly 7 $\rm \mu m$.
	We set the vibration amplitude to about 3 $\rm \mu m$ to avoid the overload of the bender.
	The amplitude corresponds to the modulation efficiency of 9.6\% (See Sec. I in Supplementary Material).

	Fig. \ref{fig:2}(b) shows the schematic of this work. The optical system is composed of a 532 nm laser, a battery of lenses, and an Acoustic Optical Modulator (AOM) to generate initialization and readout laser pulses. The microwave system consists of a microwave (MW) source, an In-phase/Quadrature (IQ) mixer, a Positive-Intrinsic-Negative diode (PIN), and a power amplifier. Microwave from the MW source is multiplied with a radio-frequency (RF) signal fed into the IQ mixer. Then the output is gated by the PIN to generate the spin-echo sequence and amplified by a power amplifier. The MW signal is finally sent to a double split-ring microwave resonator \cite{Bayat2014} to drive the NV centers.
	The Control and Readout System (CRS) shown in the figure stands for a home-built system \cite{Qin2020}. Both of the AOM and the PIN timing is controlled by the Transistor-Transistor Logic (TTL) signals from the CRS. RF signals utilized to multiply with the MW signal are generated from the Arbitrary Waveform Generator (AWG) integrated into CRS. Another waveform generator provides a sinusoid signal for the excitation of a piezoelectric bender and it is also triggered by the CRS. Inside the probe, a couple of FCs is used to amplify the external magnetic field on the diamond. The length of FC is 4 cm and the width is 8 cm. The amplified magnetic field is modulated by a vibratory modulation chip glued onto the piezoelectric bender. The fluorescence is collected through a compound parabolic concentrator contacting one side of the diamond. After being separated from the excitation laser by a Long-Pass Filter (LPF), the fluorescence is transferred to the photodiode (PD). The signal from the PD are amplified by a current amplifier and finally sampled by the CRS. Another PD, which is not shown in this figure, is utilized to cancel the fluctuation of the laser's intensity by monitoring the variation of the intensity of the laser as the reference signal \cite{Clevenson2015,Barry2017}.

	\begin{figure*}
		\includegraphics[width=1.0\textwidth]{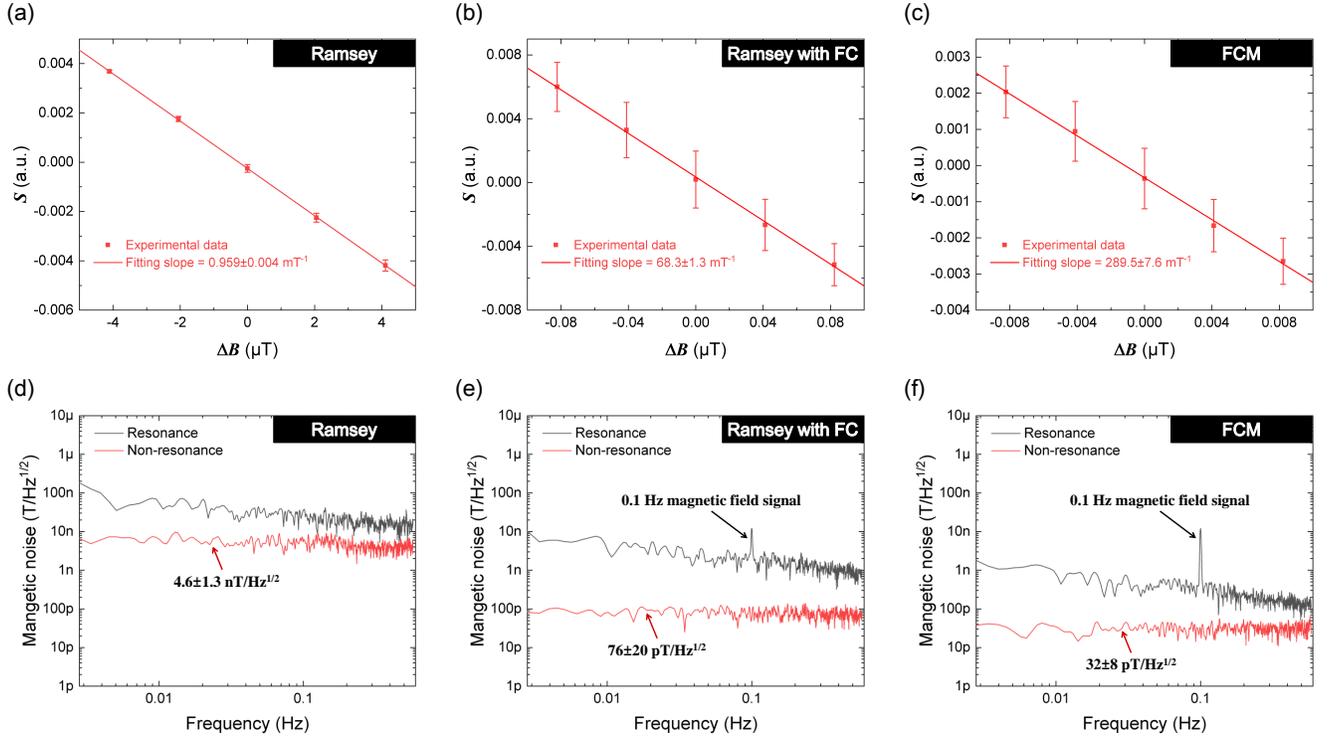}% Here is how to import EPS art
		\caption{\label{fig:3} Sensitivities of different methods. (a)-(c) The signal $S$ as the function of the applied magnetic fields for different methods. $S$ was obtained by the reference readout and the differential technique (See Sec. I in Supplementary Material). $\Delta B$ in abscissa represents for relative applied magnetic field. Four repetitive field-sweep spectra were tested. Each point on the figure was the average of four measurements. Error bars denote the standard deviation of four measurements. The maximum slope max$|\partial S/ \partial B|$ was obtained from the fitting. (d)-(f) Magnetic field amplitude spectrum density of magnetometer based on different methods. The 0.1 Hz magnetic field signal was applied by coils for calibration. Green, blue and purple dashed lines represent for sensitivities of Ramsey protocol, Ramsey protocol with FC, and FCM method, respectively.}
	\end{figure*}

	To demonstrate the enhancement of the sensitivities of NV-magnetometry, we carried out experiments to evaluate the sensitivities for three types of NV magnetometry. The first one is the Ramsey-type NV-magnetometry. The second one is Ramsey-type NV-magnetometry with FCs. Because of the magnification of the external magnetic field via FCs, the sensitivity of the second type of magnetometer is expected to be improved compared with first one without FCs. Both the sensitivities of the first and second types of magnetometry are limited by $T_2^*$. The third type is NV-magnetometry with FCM. The sensitivity of this type are expected to be further improved compared with the second one, and it is essentially limited by $T_2$. All the experiments were carried out with the same diamond in a magnetic shield.
	
	We first measured the sensitivity of Ramsey-type NV-magnetometry.
	A coil was utilized to generate the external magnetic field, which is calibrated by a Tunnel MagnetoResistance (TMR) senor (See Sec. I in Supplementary Material).
	A stepped magnetic field generated by the coil was used to test the max$|\partial S/ \partial B|$, as shown in the Fig. \ref{fig:3}(a).
	The max$|\partial S/ \partial B|$ of $0.959\pm 0.004$ mT$^{-1}$ was obtained.
	Then the time domain signal was measured by repeating the Ramsey sequence under resonance and non-resonance conditions, respectively.
	The non-resonance condition means that the microwave frequency was set to be far from the resonance frequency of the NV centers.
	The time domain signal with about one hour continuously acquisition was then converted to amplitude spectral density (ASD) \cite{Xie2021} to display the noise floor in frequency domain, as shown in the Fig. \ref{fig:3}(d).
	%Each data points was accumulated by 4000 times.(2000 for Ramsey and Rmasey with FC and 4000 for FCM)
	The sampling rate for the acquisition was 1.15 Hz with one point accumulated by 4000 times.
	The ASD was computed using the Welch's method with a 1380-point Blackman-Harris window with 50\% overlap.
	The ASD measured under resonance condition was higher than the non-resonance condition.
	%Since the estimated frequency range of down to about 0.001 Hz was low enough in ASD, the temperature fluctuation caused by laser and MW pulses becomes the main reason to separate the two ASD spectra \cite{Fescenko2020}.
	The thermal fluctuation caused by laser and MW is the the main reason to separate the two ASD spectra \cite{Fescenko2020}.
	%To avoid the impact, previous works \cite{Barry2017, Fescenko2020} reported the noise floors with the frequency range of above 100 Hz as the low-frequency sensitivity.
	To avoid this impact, previous works \cite{Jensen2014, Zheng2019, Zheng2020} using the ASDs measured under non-resonance conditions to evaluate the sensitivity of the NV-magnetometry.
	Due to the data rate limit of the acquisition system in the current setup, we met the challenge of increasing the sampling rate, so the estimated frequency range is
	from $\sim$0.003 Hz to $\sim$0.6 Hz in ASD.
	To estimate the magnetometry performance, we followed the previous works and used the non-resonant noise floor as the low-frequency sensitivity \cite{Jensen2014, Zheng2019, Zheng2020}, which was obtained as $ 4.6\pm 1.3$ nT/Hz$^{1/2}$ for Ramsey protocol in Fig. \ref{fig:3}(d).

	%Thus we used the non-resonant noise floor as the sensitivity, which was obtained as $11\pm 3$ nT/Hz$^{1/2}$ for Ramsey protocol in the Fig. \ref{fig:3}(d).
	%The spectra was obtained from 1 hour continuously acquired time domain data.
	%for most applications;disturbance;
	%, which were Ramsey protocol enhanced by FC and FCM respectively,
	
	The experimental results of sensitivity of the Ramsey-type NV-magnetometry with FCs is shown in  Fig. \ref{fig:3}(b and e).
	%It should be noticed the range of the stepped magnetic field $\Delta B$ was reduced.
	The magnification of the external magnetic field via FCs is measured to be 85.
	As shown in the Fig. \ref{fig:3}(b), the max$|\partial S/ \partial B|$ with FCs was obtained as $68.3\pm 1.3$ mT$^{-1}$, which is about two orders of magnitude of that without FCs.
	The noise floors were also obtained as shown in the Fig. \ref{fig:3}(e).
	The sensitivity of Ramsey-type NV-magnetometry with FCs is $76\pm 20$ pT/Hz$^{1/2}$.

	The experimental results for sensitivity of NV-magnetometry with FCM are shown in Fig. \ref{fig:3}(c and f).
	The vibration phase and the position of the modulation chip have been optimized (See Sec. I of Supplementary Material).
	As shown in the Fig. \ref{fig:3}(c), the max$|\partial S/ \partial B|$ of FCM method was $289.5\pm 7.6$ mT$^{-1}$.
	Due to the prolongation of $\tau$ with the FCM method, each data point was accumulated by 2000 times in order to maintain the same sampling rate of $\sim$1.15 Hz compared to the Ramsey experiment.
	The sensitivities of $32\pm 8$ pT/Hz$^{1/2}$ was obtained for FCM method.
	
	A test signal with the frequency of 0.1 Hz was applied in all sensitivity tests (See Sec. I of Supplementary Material).
	%The amplitude of this test magnetic signal was set to be 12 nT.
	In the measurement of Ramsey-type NV-magnetometry without FCs, there is no observation of such a signal. In the measurements of Ramsey-type NV-magnetometry with FCs and NV-magnetometry with FCM, there are clear observations of the test signal. It is clear that the signal to noise ratio of the test magnetic field by NV-magnetometry with FCM is much better than that from the Ramsey-type NV-magnetometry with FCs. We have successfully demonstrated the improvement of the sensitivity of dc NV-magnetometry by FCM.
	%As displayed by comparing the ASDs of resonance condition in Fig. \ref{fig:3}(e) and (f), we could conclude that a part of non-magnetic disturbances were suppressed by the FCM method.
	%These evidences proved the superiorities of the FCM method.

	%vthe magnification of the flux concentrators is 85,
	To further investigate the potential of these methods, we establish a model according to previous works \cite{Barry2020,Degen2017, Xie2021} (See Sec. II in Supplementary Material) for the sensitivity evaluation as follows,
	
	\begin{equation}\label{eq-pulse}
		\eta \approx A \frac{n_f}{G \alpha E_{_F} \gamma_e e^{-\left(\tau / T_{\rm{coh}}\right)^{p}}C\sqrt{N}} \frac{\sqrt{t_{m}+\tau}}{\tau} \rm ,
	\end{equation}
	
	\noindent where $A$ is a coefficient for different pulse sequences. $A=1$ ($A=\pi/2$) is corresponding to the Ramsey protocol (spin-echo sequence) \cite{Barry2020}. $n_f$ is the ratio between overall noise of system and the shot noise. For the shot-noise limited sensitivity, we have $n_f=1$. $G$ is the magnification of FC. $\alpha$ is the angle factor used for describing the misalignment between magnetic field and the NV symmetry axis \cite{Xie2021}. $E_{_F}$ is modulation efficiency of the FCM method (See Sec. I of Supplementary Material). $\gamma_e$ is the gyromagnetic ratio of electron. $\tau$ is the evolution time of the NV center. $T_{\rm coh}$ is coherence time of the NV centers. $p$ is stretched exponential parameter depended on the origins of the dephasing \cite{Barry2020}. %NV resonance line shape with Lorentzian profiles correspond to $p=1$ and line shapes with Gaussian profiles correspond to $p=2$.
	$C$ is the measurement contrast \cite{Taylor2008}. $N$ is the average number of photons detected per measurement. $t_{m}$ is the additional time in the pulse sequence (See Sec. II of Supplementary Material).
	%There are two important parameters in our experiment, which are G and $\tau$. For different types of NV Magnetometry demonstrated here, these two parameters varies and results in different sensitivities.

	%It should be noticed $G$ was essential for the sensitivity enhancement in this procedure.
	The parameters in the equation (\ref{eq-pulse}) can be experimentally measured as shown in table I in Supplementary Material.
	%The sensitivities of three methods were evaluated by using equation (\ref{eq-pulse}).
	%The parameters in equation (\ref{eq-pulse}) were tested,
	For the Ramsey-type NV-magnetometry without FCs, $G=1$ and $\tau=0.7~\rm \mu s$. The predicted sensitivity according to the equation (\ref{eq-pulse}) is 3.3 nT/Hz$^{1/2}$, which is consistent with the experimental result in Fig. \ref{fig:3}(d).
	For the method of Ramsey-type NV-magnetometry by FC, the predicted sensitivity by equation (\ref{eq-pulse}) is 67 pT/Hz$^{1/2}$  with the parameters $G=85$ and $\tau=0.7~\rm \mu s$. The predicted sensitivity is also consistent with the experimental result.
	For the FCM method, $\tau$ has been set to $92.7~\rm \mu s$ to match the modulation frequency.
	The spin-echo sequence prolonged the coherence time, which was $102~\rm \mu s$ in Fig. \ref{fig:1}(b).
	The experimental modulation efficiency of the FCM, $E_{_F}$, was 9.6\%. The contrast C is 0.0045.
	With $T_{\rm coh} = T_2$, the predicted sensitivity of  NV-magnetometry by FCM is 39 pT/Hz$^{1/2}$, which agrees with the experimental result.
	The slight differences between the evaluated sensitivities and the experimental values may come from the error of the parameters used in the evaluation.
	For example, the coefficient $A$ was not exactly $\pi/2$, since the modulated magnetic field was not a perfect sine.
	Nevertheless, the predicted sensitivities are consistent with the experimental results. Therefore, our model is suitable for predicting the performance of NV-magnetometry and we would like to utilize such model to evaluate the potential of sensitivity NV-magnetometry with FCM with current state-of-art technologies.
	%These evaluations give a support to the equation (\ref{eq-pulse}).
	%The sensitivity enhancement for FCM method is brought by $G$, $\tau$ and $T_{r}$.
	%Notice $E_{_F}$ decreases the sensitivity enhancement for the FCM method in the experiment.

	%Further advancement of $E_{_F}$ is feasible.
	
	%The FCM method transferred the low-frequency magnetic field into the modulation frequency of nearly 10.8 kHz.
	%the sensitivity was evaluated finally. was capable of improving the sensitivity
	
	%The key parameters were $G=100$ and $\tau=92.7 \rm \mu s$. .
	%The diamond sample with the above $T_{2}$ can have

	% %In the evaluation, the minimized $N$ of $9\times10^{9}$ is selected, which can be realized by single-pass excitation in experiment.
	%Since one may concern the sensitivity boosting was brought by flux concentration totally, we figure out the future sensitivities with improvements.
	According to detailed evaluations in the Sec. II of Supplementary Material, the sensitivity can be further improved in the following steps by cutting-edge  technologies.
	(i) When the width of the diamond is reduced, the magnification of the magnetic field by FCs can be increased.
	%while the reduction of the number of NV centers for magnetic field sensing can be avoided.
	(ii) Through polishing modulation chip and FCs together with increasing vibration amplitude of piezoelectric bender, $E_{_F}$ can be increased to above fifty percent.
	This high modulation efficiency is achievable since a recent experiment reported a modulation efficiency up to 68.7\% with a similar method \cite{Du2019}.
	%For comparison, the flux modulation used for magnetoresistive sensors has achieved the modulation efficiency up-to 68.7 \% \cite{Du2019}.
	(iii) The properties of the diamond can be further optimized. The coherence time can be improved with better diamonds together with advanced quantum control technologies\cite{Bauch2018,Wolf2015}.
	%The number of NV-centers evolved in sensing can be promoted even the width of the diamond is reduced.
	According to the detailed work on the diamond samples \cite{Barry2020}, $T_{2}$ is expected to be $\sim 700\ \rm \mu s$ with $^{12}$C enriched diamond, and the NV concentration of about 0.019 ppm is reachable.
	If the total internal reflection method \cite{Clevenson2015} is utilized,  a reliable $N$ of $2.6\times10^{10}$ with the above NV concentration is expected.
	%In our evaluation, the parameters of the samples with good quality \cite{Wolf2015, Bauch2018, Zhang2021c} are considered.
	(iv) The contrast $C$ in experiment is decreased by the inhomogeneity of the magnetic field in diamond. The inhomogeneity is mainly due to the FCs' remanence as analyzed in the Supplementary Material.
	When such remanence of FCs is reduced via the demagnetization procedure, $C$ can be promoted to $\sim1.2\times10^{-2}$.
	With above improvements, an optimized shot-noise-limited sensitivity is expected to be $\sim 3~\rm fT/Hz^{1/2}$.
	Besides, the reduction of $n_f$ is essential on the way to the shot-noise-limited sensitivity, since it is about 20 in the current experiments.
	It could be helpful to improve the laser stability and reduce the noise of data acquisition  system for decreasing $n_f$ \cite{Fescenko2020,Schloss2018,Chatzidrosos2017}.
	The further reduction of $n_f$ can lead to an achievable sensitivity of femtotesla level in the future study.
	
	%A promising route of reducing $n_f$ is to improve the laser stability and reduce the noise of data acquisition  system.
	
	%depended on the photons detected per measurement $N$ and modulation efficiency $E_{_F}$
	%There are also many parameters hard to determine, such as $n_f$, $E_{_F}$, $T_{r}$ and $N$.
	%Thus the sensitivities in a wide range are evaluated using equation (\ref{eq-pulse}), as shown in the Fig. \ref{fig:4}.

	%With the total internal reflection \cite{Clevenson2015}, the expected $N$ is $5\times10^{12}$.
	%Thus the upper bound of $N=5\times10^{12}$ in the evaluation is used, plotted as the horizontal dashed line.
	%With above improvements, an optimized shot-noise-limited sensitivity of $\sim 3~\rm fT/Hz^{1/2}$ is achievable, as the red point shown in Fig. \ref{fig:4}.

	%We also notice that the samples with better quality \cite{Wolf2015, Bauch2018, Zhang2021c} are useful references for these evaluation.
	%utilized in the sensitivity evaluation.
	%And, there are still some issues, which should be addressed for achieving better sensitivity.
	
	%The lower bound of $N$ gives the worst sensitivity of $1000~\rm fT/Hz^{1/2}$ in the Fig. \ref{fig:4}.
	%The figure gives a comparison for the future optimizations.
	
	Our work demonstrates a dc quantum magnetometry with $T_2$-limited based on diamonds via flux modulation.
	%The limitations for enhancement of the sensitivity with the current magnetometry have been  systematically analyzed.
	%With the state-of-art technologies, the sensitivity can be further improved to femtotesla level.
	Compared to the recent progress in dc NV-magnetometry via rotating the diamond, our method have a unique advantage as follows. Since FCM can be realized by Micro-ElectroMechanical Systems (MEMS),  the power consumption and the size of the sensor can be further reduced. Very recently, a CMOS-integrated NV-Magnetometry has been successfully demonstrated \cite{Kim2019}. We further anticipate that a CMOS-integrated NV-Magnetometry with FCM by MEMS will provide a highly integrated low frequency magnetometry with high sensitivity.
	Our work provides a clear route towards femtotesla magnetometry at room temperature, which will play an important role in  biological signal detection, such as magnetocardiography.

	\section*{Acknowledgments}
	We thank Wenzhe Zhang for the helpful discussion.
	This work was supported by the Chinese Academy of Sciences (Grants No. XDC07000000, No. GJJSTD20200001, No. QYZDY-SSW-SLH004, No. QYZDB-SSW-SLH005), Innovation Program for Quantum Science and Technology (Grant No. 2021ZD0302200), the National Key R$\&$D Program of China (Grant No. 2018YFA0306600), the National Natural Science Foundation of China (Grant No. 81788101), Anhui Initiative in Quantum Information Technologies (Grant No. AHY050000), Hefei Comprehensive National Science Center, and the Fundamental Research Funds for the Central Universities. X. R thank the Youth Innovation Promotion Association of Chinese Academy of Sciences for the support.
	
	%\textbf{Author contributions:}
	%J.D. and X.R. proposed the idea and supervised the experiments. Y.X., H.Y., Y.Z. and X.Q. prepared the experimental setups. Y.X. and Y.Z. performed the experiments. H.Y. prepared the sample. Y.Z. and C.K.D. performed the simulations. X.R., Y.X., H.Y. and Y.Z wrote the paper. All authors analysed the data, discussed the results and commented on the manuscript.
	%\textbf{Competing interests:}
	%The authors declare that there are no competing interests.
	%\textbf{Data and materials availability:}
	%All data are available in the manuscript or the Supplementary Material.
	
	Y. X., C. X. and Y. Z. contributed equally to this work.

%	\bibliographystyle{Science}
%	\bibliography{ref, Supplementarymaterial}% Produces the bibliography via BibTeX.

	%\begin{addendum}
	%\item [Acknowledgements]
	%This work was supported by the Chinese Academy of Sciences (Grants No. XDC07000000, No. GJJSTD20200001, No. QYZDY-SSW-SLH004, No. QYZDB-SSW-SLH005), the National Key R$\&$D Program of China (Grant No. 2018YFA0306600 and No. 2016YFB0501603), the National Natural Science Foundation of China (Grant No. 81788101), and Anhui Initiative in Quantum Information Technologies (Grant No. AHY050000). X. R thank the Youth Innovation Promotion Association of Chinese Academy of Sciences for the support.
	%
	%
	%\item[Author contributions]
	%J. D. and X. R. proposed the idea and supervised the experiments. Y. Z., Y. X. and K. J. designed and performed the experiments. H. Y. prepared the sample. W. Z. and X. Q. provided the home-built lock-in amplifier. Y. Z., Y. X. and K. J.  carried out the simulations. All authors contributed to the preparation of the manuscript. All authors analysed the data, discussed the results and commented on the manuscript.
	%
	%\item[Competing Interests] The authors declare that there are no competing interests.
	%
	%\item[Additional information]
	%Correspondence and requests for materials should be
	%addressed to J.D. (djf@ustc.edu.cn) and X. R. (xrong@ustc.edu.cn).
	%\end{addendum}
	
	%%
	%% TABLES
	%%
	%% If there are any tables, put them here.
	%%
	\clearpage
%	\onecolumngrid
	
	\vspace{1.5cm}
	\begin{center}
		\textbf{\large Supplementary Material}
	\end{center}

	\appendix
	\tableofcontents
	\renewcommand{\thefigure}{S\arabic{figure}}
	\setcounter{figure}{0}
	\renewcommand\thesection{\Roman{section}}
	\section{Materials and methods}

	\subsection{Sample preparation}
	
	The diamond sample used in this work was a single crystal chip (Element Six, DNV B1), grown by using chemical vapor deposition (CVD). Its typical initial nitrogen concentration and NV concentration were 0.8 ppm and 0.3 ppm, respectively. The diamond sample was cut into a piece with the size of 1.0 mm $\times$ 1.0 mm $\times$ 0.4 mm, and the 1.0 mm $\times$ 1.0 mm facet is perpendicular to $[100]$ crystal axis.
	
	\subsection{Probe with flux concentration and modulation}
	
\begin{figure*}[!htp]
		\includegraphics[width=1.0\textwidth]{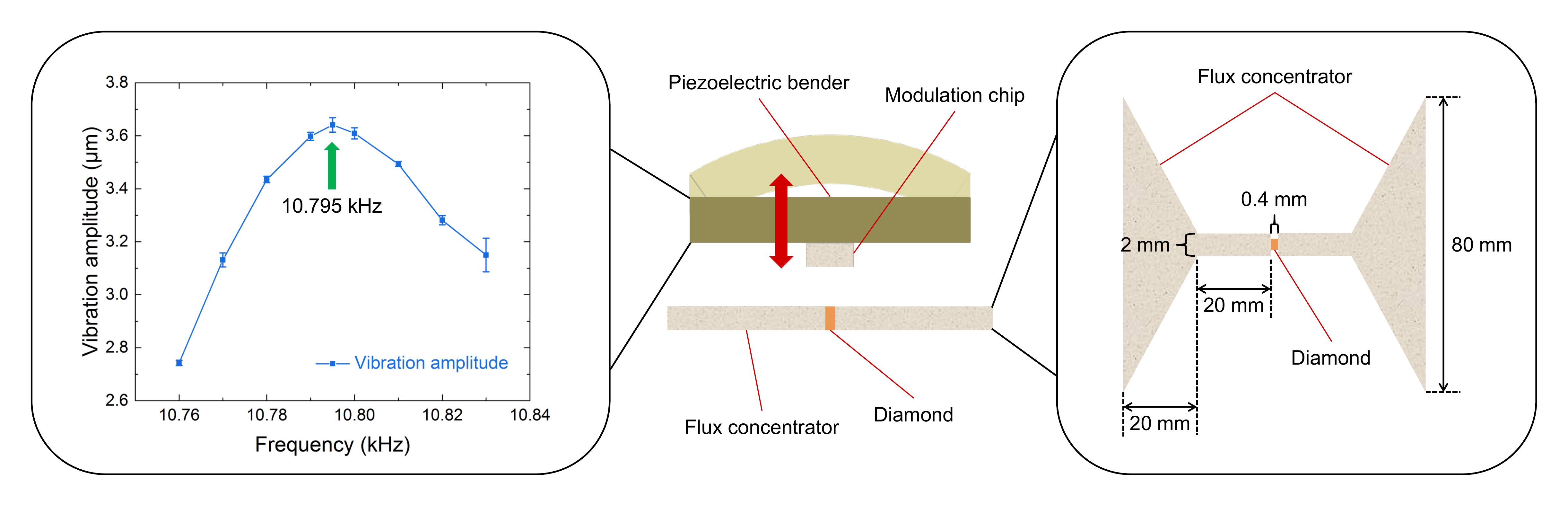}% Here is how to import EPS art
		\caption{\label{fig:S1}Schematic of probe with FCM. The red double headed arrow indicates the vibration direction of piezoelectric bender. Black box on left side shows frequency-response curve of the piezoelectric bender. The frequency-response curve was obtained with the load of modulation chip. Black box on right side shows the size of FCs.}
	\end{figure*}

	As shown in Fig.~\ref{fig:S1}, the probe with Flux Concentration and Modulation (FCM) contains a commercial piezoelectric bender (Harbin Core Tomorrow, NAC2223), a modulation chip and two Flux Concentrators (FCs).
	The piezoelectric bender is a cuboid piezo-bimorph, whose size is 21.0 mm $\times$ 7.8 mm $\times$ 1.8 mm.
	Both end of the piezoelectric bender are glued onto an aluminium alloy holder.
	The modulation chip is a cuboid made of 1J85 alloy, whose size is 4 mm $\times$ 2 mm $\times$ 1 mm.
	The modulation chip is glued onto the center of piezoelectric bender.
	The size of the FCs is shown in Fig.~\ref{fig:S1}.
	The FCs are specially shaped sheet metals made of 1J85 alloy.
	The thickness of FC is 1 mm.
	
	The resonance frequency and the corresponding vibration amplitude of the piezoelectric bender were 10.795 kHz and 3.6 $\mu$m with the excitation voltage of 75 V, as shown in Fig.~\ref{fig:S1}.
	In sensitivity tests, to protect the piezoelectric bender, the vibration amplitude was set to about 3.0 $\mu$m with the excitation voltage of 30 V.
	The vibration amplitude was tested by a high speed laser doppler vibrometer (Sunny Optical, LV-S01).

	%\subsection{FC installation}
	
	%Height difference between the end facet of two FCs was a major factor that affects the modulation efficiency. Due to simulation using finite element method software (COMSOL Multiphysics 5.6), a height difference of $30\,\mu m$ can lead to more than 6 times decrease in modulation efficiency. To solve this problem, micropositioning stage and position sensor were used in this work to accurately locate FCs. During the installation process, FCs were connected to micropositioning stage for precise moving. Height difference was real-time measured by position sensor for micropositioning stage control feedback. With several adjustments, the height difference could be reduced to less than $5\,\mu m$ and modulation efficiency could be enhanced by 5 times over traditional FC installation method.
	
	\subsection{Position optimization of the modulation chip}

	\begin{figure*}[!htp]
		\includegraphics[width=0.9\textwidth]{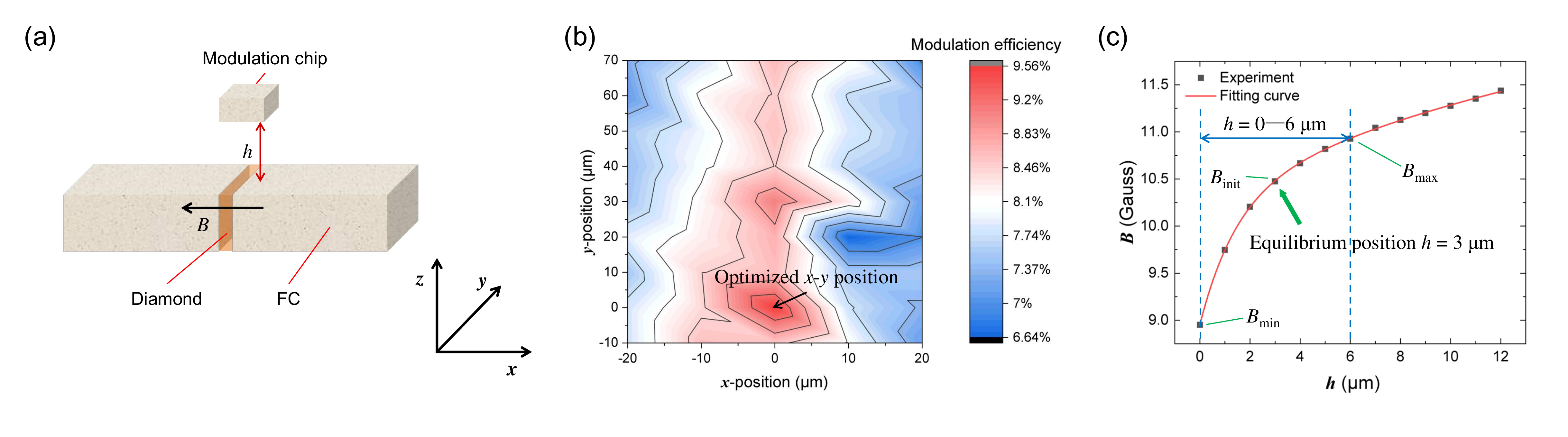}% Here is how to import EPS art
		\caption{\label{fig:S2}Position optimization of the modulation chip. (a) Schematic of the modulation chip and the coordinate system. $h$ is the the minimum distance between modulation chip's lower surface and FCs' upper surface. $h=0$ means the two surfaces are in contact. (b) The \textit{x-y} position optimization of the modulation chip. An optimal \textit{x-y} position is indicated by the arrow. (c) $B$ as the function of $h$ measured at the optimal \textit{x-y} position.}
	\end{figure*}
	
	%	The width of gap $h$ is the minimum distance between modulation chip's lower surface and FCs' upper surface, denoted in Fig.~\ref{fig:S2}(a).
	%	$B$ is the magnetic field in diamond between FCs, which is obtained by measuring the CW spectra.
	%	The Continuous Wave (CW) spectra measured under different $h$. The resonance frequency extracted from the CW spectra is used to calculate the $B$, the magnetic field in diamond between FCs. The angle between the direction of $B$ and the axis of NV centers is 54.7 degree. Vertical shifts is added manually for clear view. The triangle points on each spectral line represent for resonance frequencies. (c)
	
	The position of the modulation chip should be optimized to achieve high modulation efficiency. 	
	The modulation efficiency $E_{_F}$, which is an important parameter for the sensitivity of FCM method, can be defined as \cite{pan12}
	
	\begin{equation}\label{func1}
		E_{_F}=\frac{B_{\rm max}-B_{\rm min}}{2B_{\rm init}} \times 100\% .
	\end{equation}
	
	$B_{\rm max}$ and $B_{\rm min}$ are maximum and minimum intensity of magnetic field in diamond clamped by FCs during the flux modulation. $B_{\rm init}$ is the intensity of magnetic field in diamond between FCs with modulation chip at the equilibrium position.
	
	According to the equation (\ref{func1}), the optimized \textit{x-y} position of the modulation chip can be achieved by changing $h$ and measuring $B$ at different \textit{x-y} positions, denoted as \textit{B-h} curve.
	The schematic of the FCM method and the coordinate system is shown in Fig.~\ref{fig:S2}(a).
	%	The modulation efficiency at specified \textit{x-y} position can be calculated by (\ref{func1}) via \textit{B-h} curve.
	A magnetic field of about 12.3 $\mu$T is applied during the optimization process.
	%	Fig.~\ref{fig:S2}(b) shows the CW spectra examples measured with $h=$ 0, 1, 2, 3 $\mu$m.
	%	Significant variation of resonance microwave frequency can be seen.
	The magnetic field $B$ in diamond is obtained by measuring the resonance frequency extracted from the CW spectra.
	The optimization of the modulation chip's \textit{x-y} position is shown in Fig.~\ref{fig:S2}(b).
	The irregular shape in Fig.~\ref{fig:S2}(b) is due to the imperfect surfaces of the modulation chip and the FCs, discussied in Sec. \ref{subsec-EF}.
	Fig.~\ref{fig:S2}(c) shows the \textit{B-h} curve measured at the optimal \textit{x-y} position shown in Fig.~\ref{fig:S2}(b).
	$B$ decreases with $h$ decreasing.
	Since the vibration amplitude of piezoelectric bender is set to 3 $\mu$m, the equilibrium position of modulation chip is chosen as $h=3~\mu$m to achieve maximal modulation efficiency.
	$B_{\rm max}=$ 10.95 Gauss, $B_{\rm min}=$ 8.95 Gauss and $B_{\rm init}=$ 10.47 Gauss with above configuration are indicated on the Fig.~\ref{fig:S2}(c).
	Thus $E_{_F}=9.6$\% is obtained by using (\ref{func1}).
	%	With the optimization, the modulation efficiency of about 9.6$\%$ was realized.

	%	It should be noticed that the surfaces of the modulation chip and the FCs are rough.
	%	Thus, $h=0$ means the contact between the modulation chip's lower surface and the FCs' upper surface.

	%	$B$ represents for magnetic field in interval between FCs, which was calculated by resonance microwave frequency. $h$ is the width of gap between modulation chip and FCs. Dashed boxes represent for the modulation chip moved to different positions during optimization process.
	
	\subsection{Vibration phase optimization of the piezoelectric bender}
	%	The dynamic phase accumulation of ;;	Accumulated dynamic phase and frequency of fringe on ;;when the phase match the $\pi/2$-pulse
	%Due to the unknown vibration delay and modulation delay time, phase of piezoelectric bender vibration should be optimized.
	%	shows the spin-echo sequence.  Signal to magnetic field response were measured under different advance times during the optimization process.
	\begin{figure*}[!htp]
		\includegraphics[width=0.9\textwidth]{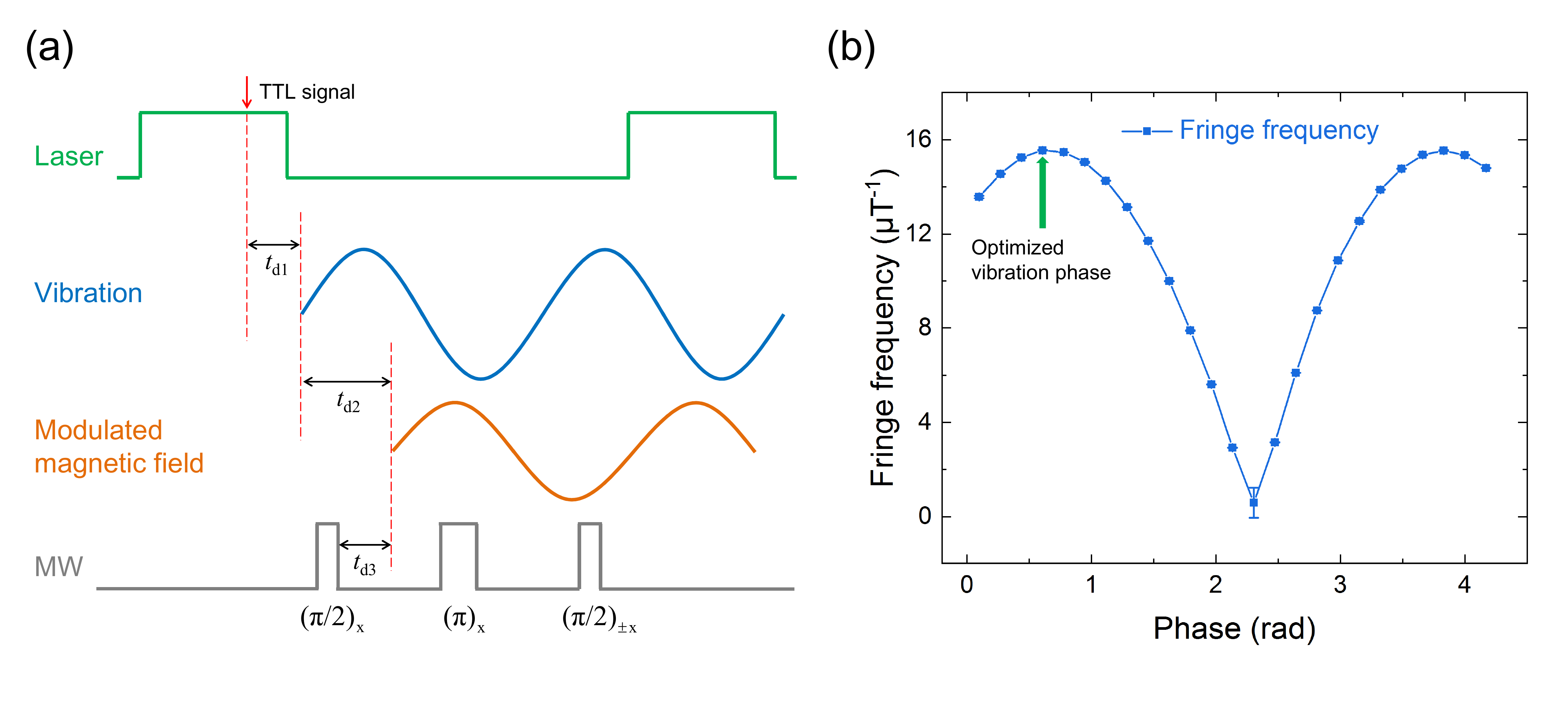}% Here is how to import EPS art
		\caption{\label{fig:S3} Vibration phase optimization of the piezoelectric bender. (a) Spin-echo sequence and modulated magnetic field. (b) The frequency of the fringe as the function of the relative vibration phase. The phase is calculated by the trigger time of TTL signal. The optimal vibration phase is indicated by green arrow on the figure. The frequency of the fringe is obtained by fitting with sine function.}
	\end{figure*}

	Fig.~\ref{fig:S3}(a) defines three delay times.
	$t_{d1}$ is the delay time between a Transistor-Transistor Logic (TTL) signal and the beginning of piezoelectric bender vibration.
	$t_{d2}$ is the delay time between the beginning of piezoelectric bender vibration and the equilibrium point of the modulated magnetic field. The equilibrium point is defined as the intensity of the modulated magnetic field equivalent to $B_{\rm init}$.
	$t_{d3}$ is the delay time between the equilibrium point of the modulated magnetic field and the ending of the first $\pi/2$-pulse in the spin-echo sequence.
	The vibration phase of the piezoelectric bender is depended on the three delay times, compared to the first $\pi/2$-pulse in the spin-echo sequence.
	
	%	The was used to change the delay time, as shown in Fig.~\ref{fig:S3}(a).
	%Vibration delay time represents for time interval between the TTL signal and the beginning of piezoelectric bender vibration. Modulation delay time represents for the delay time between the equilibrium point of the modulated magnetic field and the beginning of the piezoelectric bender's vibration.	
	$t_{d3}$ is essential for the spin-echo magnetometry.
	Since the three delay times are fixed, we optimize $t_{d3}$ by changing the trigger time of the TTL signal.
	To obtained the optimal sensitivity, we measured the fringe of the signal $S$ as the function of an applied static magnetic field under the different TTL signal delay, and calculated the frequency of the fringe.
	In experiment, a TTL signal was sent to waveform generator to trigger piezoelectric bender's vibration.
	Fig.~\ref{fig:S3}(b) shows the vibration phase optimization of piezoelectric bender.
	The abscissa was calculated from the trigger time divided the vibration period.
	The optimized trigger time for an optimal vibration phase was obtained with the optimization.

	\subsection{Experimental setup}
	
	The 532-nm laser was provided by a high-power optically pumped semiconductor laser (Coherent, Verdi G5). The microwave was generated by a microwave frequency synthesizer (National Instrument, FSW-0010). The IQ mixer (Marki, MLIQ0218L), the PIN (Mini-Circuits, ZASWA-2-50DRA+), the customized 20 W power amplifier and the double split-ring microwave resonator were used for microwave pulse sequence generation. The piezoelectric bender was driven by a waveform generator (RIGOL, DG812). The home-built CRS based on Field-Programmable Gate Array (FPGA) was used to generate TTL signals to AOM (ISOMET, M1133-aQ80L-1.5), PIN and the waveform generator. The RF modulation signal, generated by CRS, was applied to IQ mixer. The fluorescence was detected by a PD (Thorlabs, SM05PD1A) after collected by a compound parabolic concentrator (Edmund Optics, \#65-441). The photocurrent was amplified by a current amplifier (FEMTO, DHPCA-100) and finally sampled by the CRS.

	\subsection{Experimental method}
	
	The external magnetic field was applied by a coil placed near the probe. The coil's conversion coefficient of 4.1 $\mu$T/V was obtained using a Tunnel Magnetoresistance (TMR) magnetometer (MultiDimension Technology, USB27053). The 0.1 Hz test signal in sensitivity measurements was applied by another coil via the 0.1 Hz alternating voltage fed into it.
	The intensity of the test signal was 12 nT.
	The probe and coils for applying magnetic field were placed in a magnetic shield.
	All the experiments were performed under the magnetic shielding condition.
	
	%	The NV coherence time ($T_{2}^{\ast}$ and $T_{2}$), maximum slope in signal to magnetic field response and sensitivity were measured under applied magnetic field with strength near zero to reduce magnetic field inhomogeneity's impact on signal.
	\begin{figure*}[!htp]
		\includegraphics[width=0.9\textwidth]{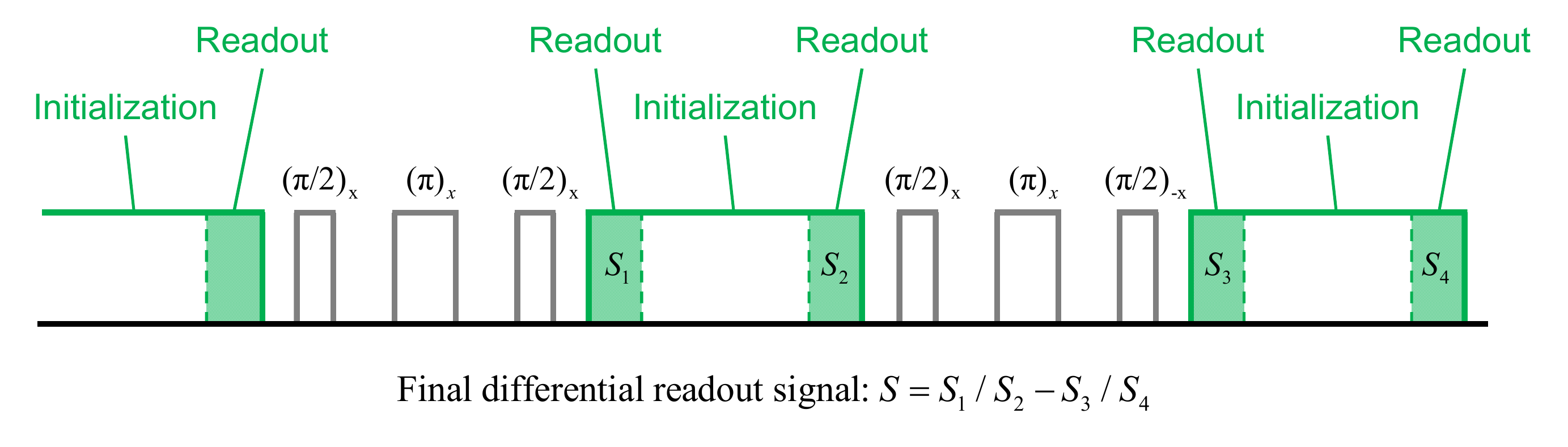}%
		\caption{\label{fig:S4}Spin-echo sequence and the readout procedure used in this work. $S_{1}$, $S_{2}$, $S_{3}$ and $S_{4}$ represents the integrated signal in the specified readout duration, respectively. The final differential readout signal $S$ can be obtained as $S=S_{1}/S_{2}-S_{3}/S_{4}$.}
	\end{figure*}

	The readout procedure used in this work are shown by Fig.~\ref{fig:S4}.
	The laser pulse in the spin-echo sequence contained the initialization duration and the readout duration.
	The intensity of the optical signal was recorded in readout duration.
	%	Photocurrent was acquired in readout duration for quantum state readout.
	A laser pulse contained one initialization duration and two readout durations.
	Data acquired in the first readout duration was divided by data acquired in second readout duration for normalization.
	
	To suppress common mode noise, another differential method was used.
	In experiment, the first microwave pulse sequence $(\pi/2)_{x}-(\pi)_{x}-(\pi/2)_{x}$ and the second microwave pulse sequence $(\pi/2)_{x}-(\pi)_{x}-(\pi/2)_{-x}$ were played alternately.
	The second microwave pulse sequence was realized by reversing the phase of the second $\pi/2$ pulse.
	Signal acquired from the first microwave pulse sequence was subtracted by the signal of the second microwave pulse sequence to suppress the common mode noise.

	\section{Further sensitivity improvement}
	
	\subsection{Magnetic field sensitivity}
	
	%	Magnetic field sensitivity is defined as \cite{jensen14,xie21}
	%	
	%	\begin{equation}
	%		\eta_{mea}=\delta S\sqrt{T_{m}}/{\rm max}|\partial S/\partial B|,
	%	\end{equation}
	
	The shot-noise-limited magnetic field sensitivity of the spin-echo protocol can be defined as \cite{Barry2020,Degen2017}
	
	\begin{equation}\label{eq1}
		\eta_{\rm spin-echo}^{\rm shot} \approx \frac{\pi}{2} \frac{\hbar}{\Delta m_{s} g_{e} \mu_{B}}  \frac{1}{e^{-\left(\tau / T_{2}\right)^{p}}C\sqrt{N}}  \frac{\sqrt{t_{I}+\tau+t_{r}}}{\tau}
	\end{equation}
	
	\noindent where $\pi/2$ is corresponding to the spin-echo protocol \cite{Barry2020}. $\hbar$ is reduced Planck constant. $\Delta m_s$ is the difference in spin quantum number between the two interferometry states. $g_e$ is the Landég-factor of electron spin. $\mu_B$ is Bohr magneton. $\tau$ is the evolution time of the NV center. $T_{2}$ is coherence time of the NV center. $p$ is stretched exponential parameter depended on the origins of the dephasing \cite{Barry2020}. $C$ is the measurement contrast \cite{Barry2020}. $N$ is the average number of photons detected per measurement. $t_I$ and $t_r$ are the duration of initialization and readout respectively.
	
	%	\noindent where $\delta S$ is minimum detectable signal intensity measured under non-resonant condition. ${\rm max}|\partial S/\partial B|$ is the maximum slope in signal to magnetic field response. $T_{m}=\tau+t_{m}$ is the total time consumption per measurement. $\tau$ is evolution time and $t_{m}$ is the sum of initialization time and readout time.
	
	We set $\gamma_e={\hbar}/{g_{e} \mu_{B}}$ and $t_{m}=t_r+t_I+t_d$ for simplifying the equation (\ref{eq1}). $t_d$ is the multiple time delays in pulse sequence with minor contribution to $t_{m}$.
	To match different types of NV-magnetometry, the coefficient ${\pi}/{2}$ is replaced by $A$, and the coherence time $T_{2}$ is replaced by $T_{\rm coh}$.
	$A=1$ ($A=\pi/2$) is corresponding to the Ramsey protocol (spin-echo protocol) \cite{Barry2020}.
	According to the previous work of the flux concentration \cite{Xie2021}, we add the the magnification $G$ of FC and the angle factor $\alpha$.
	The modulation of the magnetic field introduces a new parameter $E_{_F}$ \cite{pan12}, defined in the equation (\ref{func1}).
	For the Ramsey-type magnetometry, we have $E_{_F}=1$ since the external magnetic field is not modulated.
	Because the system noise must be larger than the shot noise, we define $n_f$ as the ratio between overall system noise and the shot noise.
	%	For the three methods (Ramsey protocol, Ramsey with FC and FCM method) used to detect magnetic field in this work, the sensitivity can be evaluated by \cite{degen17,xie21,barry20,pham13}
	With the above preconditions, the equation (\ref{eq1}) can be rewritten as

	%	\begin{equation}
	%		\begin{aligned}
	%			\eta_{eva}&=\frac{\delta S\sqrt{t_{m}+\tau}}{e^{-(\tau/T_{c})^p}C{\rm max}|\frac{\partial\cos(\frac{1}{A}G\alpha E_{_F}\gamma_{e} B\tau)}{\partial B}|}\\
	%			&=A\frac{\delta S}{G\alpha E_{_F}\gamma_{e}e^{-(\tau/T_{c})^p}C}\frac{\sqrt{t_{m}+\tau}}{\tau},\label{func2}
	%		\end{aligned}
	%	\end{equation}
	
	\begin{equation}\label{eq-pulse}
		\eta \approx A \frac{n_f}{G \alpha E_{_F} \gamma_e e^{-\left(\tau / T_{\rm coh}\right)^{p}}C\sqrt{N}} 	\frac{\sqrt{t_{m}+\tau}}{\tau} \rm ,
	\end{equation}
	
	\noindent where $\alpha$ is the angle factor used for describing the misalignment between magnetic field and the NV symmetry axis \cite{Xie2021}. For the shot-noise limited sensitivity, we have $n_f=1$. $\gamma_e$ is the gyromagnetic ratio of electron. $T_{\rm coh}$ is coherence time of the NV centers.
	
	%	There are two important parameters in our experiment, which are G and $\tau$. For different types of NV Magnetometry demonstrated here, these two parameters varies and results in different sensitivities.

	%	The $\delta S$ can be evaluated by
	%	
	%	\begin{equation}
	%		\delta S\approx \frac{n_{f}}{\sqrt{N}},\label{func3}
	%	\end{equation}
	%	
	%	\noindent where $n_{f}$ is the ratio between overall noise of system and shot noise. $N$ is average number of photons detected per measurement. By putting (\ref{func3}) into (\ref{func2}), the function can be finally written as
	
	%	\begin{equation}
	%		\eta_{eva}\approx A\frac{n_{f}}{G\alpha E_{_F}\gamma_{e}e^{-(\tau/T_{c})^p}C\sqrt{N}}\frac{\sqrt{t_{m}+\tau}}{\tau}.\label{func4}
	%	\end{equation}
	
	According to the experimental data, the parameters of three methods in equation (\ref{eq-pulse}) are shown in Table~\ref{tab:table1}. The $p=1$ is used for the Ramsey-type magnetometry \cite{Barry2020}. From the equation (\ref{eq-pulse}), the duty cycle ${{\tau}}/({t_{m}+\tau})$ is also important for the sensitivity enhancement. From Table~\ref{tab:table1}, we know the duty cycle increase from $<1\%$ to $\sim40\%$. The NV-magnetometry with FCM gives a sensitivity enhancement of 140 folds compared to the Ramsey-type NV-magnetometry.
	%	The decrease of $C$ is caused by the magnetic field inhomogeneity, which will be discussed later.
	%	$\tau$ is optimized by measuring the signal as the function of the applied magnetic field.
	%	The main parameters that affect the sensitivity are $T_{\rm coh}$, $N$, $G$, $E$ and $C$. Works need to be done on optimizing these parameters, for further improvement on sensitivity of FCM method, will be discussed in next chapters.
	%	\footnote{}
	
	%\footnote{The noise of piezoelectric bender's vibration amplitude, frequency and phase lead to increase of noise.}
	
	\begin{table*}
		\caption{\label{tab:table1}Parameters used in the sensitivity evaluations}
		\begin{ruledtabular}
			\begin{tabular}{ccccc}
				Parameter&Ramsey&Ramsey with FC&FCM\\ \hline
				$A$&1&1&$\pi$/2\\
				$G$&1&85.1&85.1\\
				$E_{_F}$&1&1&0.096\\
				$N$&7.6 $\times$ 10$^{9}$&7.0 $\times$ 10$^{9}$&7.0 $\times$ 10$^{9}$\\
				$C$&1.2 $\times$ 10$^{-2}$ &9.2 $\times$ 10$^{-3}$&4.5 $\times$ 10$^{-3}$\\
				$T_{\rm coh}$&1.13 $\mu$s&1.13 $\mu$s&102 $\mu$s\\
				$\tau$&0.7 $\mu$s&0.7 $\mu$s&92.7 $\mu$s\\
				$t_{m}$&115 $\mu$s&115 $\mu$s&140 $\mu$s\\
				$n_{f}$&12.2&15.6&19.2\\
				$p$&1&1&1.24\\
				$\gamma_{e}$&2$\pi$ $\times$ 28 GHz/T&2$\pi$ $\times$ 28 GHz/T&2$\pi$ $\times$ 28 GHz/T\\
				$\alpha$&0.5774&0.5774&0.5774\\ \hline
				$\eta$ (Experiment)&4.6 nT/Hz$^{1/2}$&76 pT/Hz$^{1/2}$&32 pT/Hz$^{1/2}$\\
				$\eta$ (Evaluation)&3.3 nT/Hz$^{1/2}$&67 pT/Hz$^{1/2}$&39 pT/Hz$^{1/2}$\\
			\end{tabular}
		\end{ruledtabular}
	\end{table*}

	\subsection{Improvement of the magnification ($G$)}\label{sec-G}
	%	(a) The magnification as the function of the gap width $d$ without the modulation chip nearby. Green dashed lines indicate magnification with 0.4 mm and 40 $\mu$m interval between FCs. (b)
	% to match the improved vibration amplitude in the future
	%	imperfect modulation chip
	%	The lower magnification $G$ compared to the experiment result in table \label{tab:table1} due to the .
	\begin{figure*}[!htp]
		\includegraphics[width=0.5\textwidth]{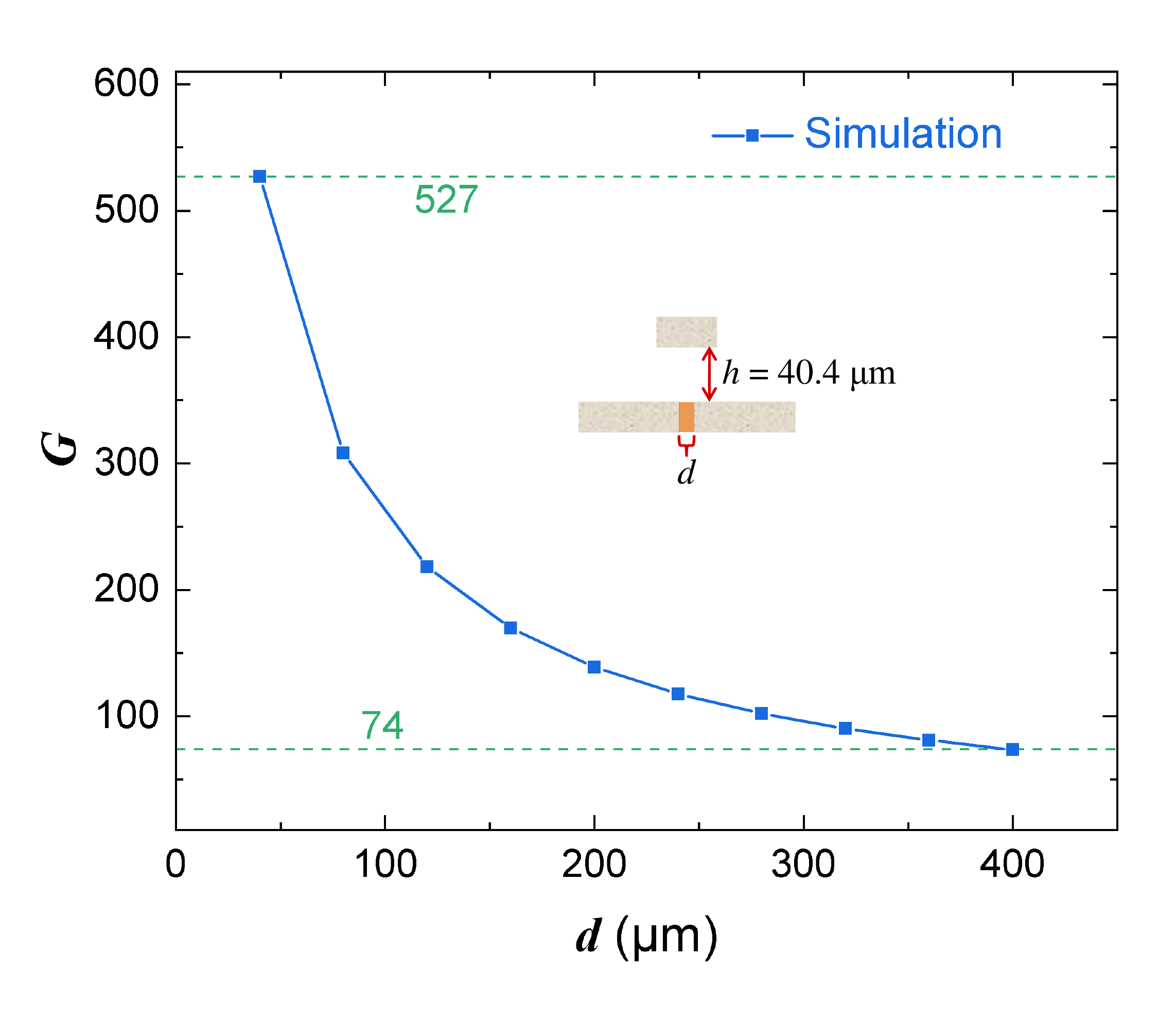}% Here is how to import EPS art
		\caption{\label{fig:S5} The simulation magnification as the function of the gap width $d$ with the modulation chip nearby. Noticed that a modulation chip exists in the simulation. $h$ is set to 40.4 $\mu$m. The impact of the imperfect modulation chip will be discussed in the section \ref{subsec-EF}.}
	\end{figure*}
	%	The FCs used in this work were made of 1J85 alloy whose relative permeability is about 30000 \cite{Xie2021}.
	%	It is difficult to find a common material with higher permeability and improvement on FC permeability can be completely limited.
	%	Geometry of FC includes shape and size of FC and interval between FCs. Since the shape and size of FC used in this work has been optimized, the only parameter can be improved is interval between FCs.
	The magnification $G$ is decided by the geometry of FCs and the gap width between FCs \cite{Xie2021}. 
	Considering the future applications in biomagnetism, we only optimize the gap between FCs to increase $G$.
	In experiment, the gap width is 0.4 mm, which is the thickness of the diamond sample.
	Diameter of the laser spot on diamond is about 40 $\mu$m.
	So the minimum gap width between FCs can be reduced to 40 $\mu$m by thinning diamond sample.
	
	The simulation magnification as the function of the gap width $d$ are shown in Fig.~\ref{fig:S5}.
	The magnification of the FCs increases rapidly with the decrease of $d$.
	According to the simulation result, the magnification increases from 74 to 527 with the gap width reducing from 0.4 mm to 40 $\mu$m.
	%	No matter whether there is a modulation chip nearby, , which represent the simulation results with and without the modulation chip nearby
	
	%	The simulation result is shown in Fig
	%	of practical situation that modulation chip is positioned right above FCs

	\subsection{Improvement of the modulation efficiency ($E_{_F}$)}\label{subsec-EF}

	\begin{figure*}[!htp]
		\includegraphics[width=0.7\textwidth]{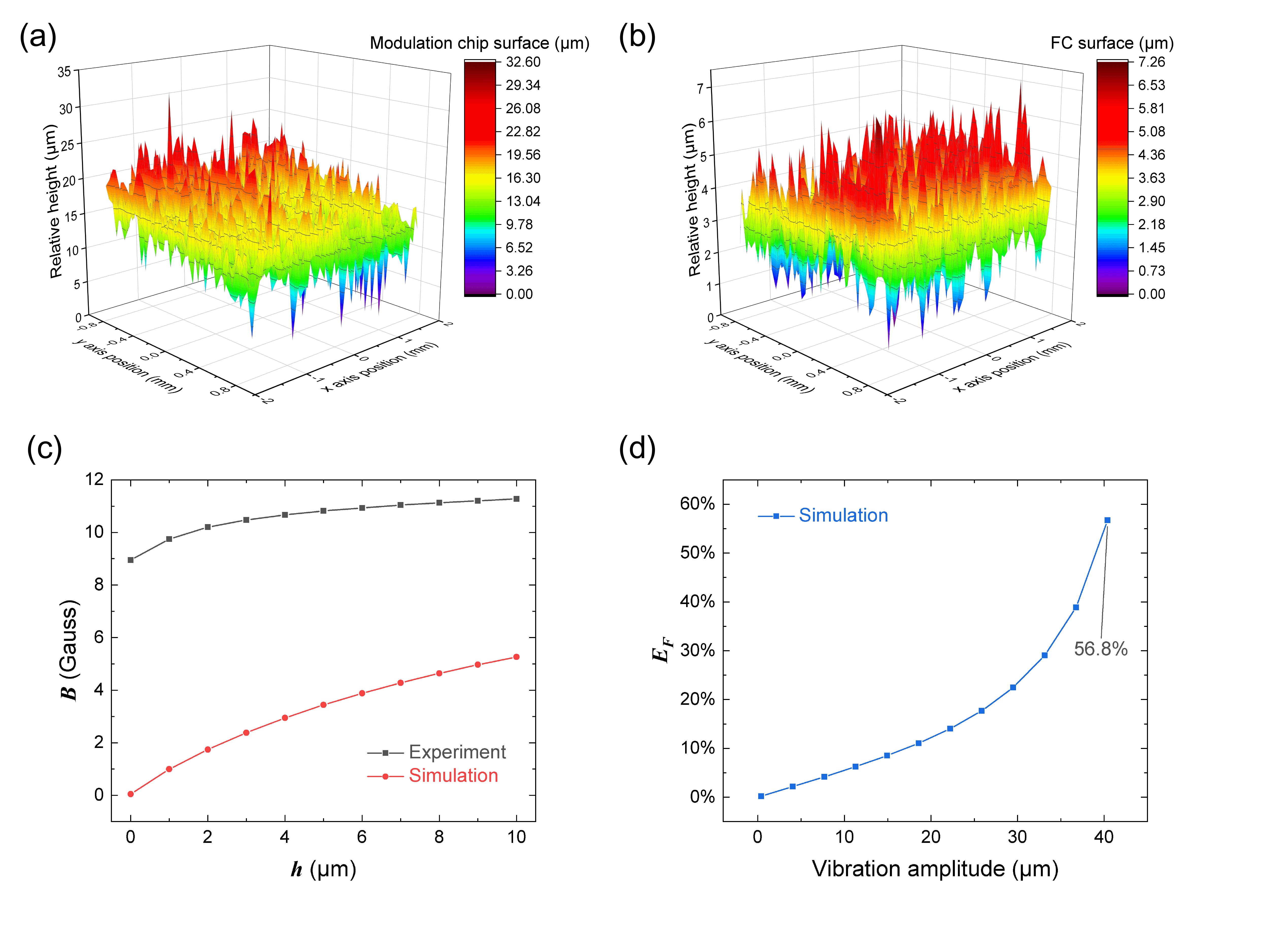}% Here is how to import EPS art
		\caption{\label{fig:S6} The surface morphology and the modulation efficiency. (a)-(b) The surface morphology of the modulation chip and the FCs, respectively. The relative height is measured by a position sensor and subtracted by minimum relative height for a clear view. (c) The magnetic field in diamond versus $h$. The simulation is conducted by assuming the ideal configuration of the modulation chip and the FCs, which are perfectly parallel and smooth. (d) The modulation efficiency as the function of the vibration amplitude. The equilibrium position of the modulation chip is set as $h=$ 40.4 $\mu$m in the simulation.}
	\end{figure*}
	%	Significant decrease of modulation efficiency can be seen.
	
	%	The nonzero dihedral angle and the surface roughness will lead to an extra gap compared to the ideal condition.
	%	The extra gap will lead to attenuation in modulation efficiency. %	If and only if the gap is thin enough that modulation chip has significant impact on magnetic field in interval between FCs. 	
	%the dihedral angle between surface of modulation chip and FCs is nonzero due to the assembly error.
	%Besides, the surfaces of modulation chip and FCs are rough.
	$E_{_F}$ is affected by the gap width between the modulation chip and the FCs.
	In experiment, the surfaces of modulation chip and FCs are rough.
	The surface roughness  of the modulation chip and the FCs are shown in Fig.~\ref{fig:S6}(a) and Fig.~\ref{fig:S6}(b). According to the figures, the surface roughness of modulation chip and FC are larger than 30 $\mu$mRz and 7 $\mu$mRz, respectively.
	In experiment, the width of gap $h$ is the minimum distance between modulation chip's lower surface and FCs' upper surface.
	Fig.~\ref{fig:S6}(c) shows the results of the magnetic field in diamond between FCs as the function of $h$.
	The simulation result in Fig.~\ref{fig:S6}(c) is an ideal configuration of the modulation chip and the FCs.
	It is noted that $h=0$ (simulation) corresponds to the perfect contact of the modulation chip and the FCs, which leads to the magnetic field intensity in diamond of nearly zero.
	
	The other key parameter, which has the impact on the modulation efficiency, is the vibration amplitude of piezoelectric bender.
	$B_{\rm max}$ and $B_{\rm min}$ in (\ref{func1}) are decided by vibration amplitude as shown in Fig.~\ref{fig:S2}(d). Fig.~\ref{fig:S6}(d) shows the modulation efficiency as the function of the vibration amplitude.
	In Fig.~\ref{fig:S6}(d), the equilibrium position of modulation chip is set as $h=$ 40.4 $\mu$m.
	With the vibration amplitude reaching 40.4 $\mu$m, the modulation efficiency of 56.8$\%$ can be achieved.
	%	Modulation efficiency increased rapidly with vibration amplitude.

	%	Extra gap induced by surface roughness can be reduced by precise polishing.
	%	Except for traditional mechanical polishing method, using  for polishing were realized and roughness of several nanometers was achieved.
	%	The improvement of the piezoelectric bender's vibration amplitude can be achieved by changing the material of the piezoelectric bender and optimizing the geometry of the piezoelectric bender.
	%	The vibration amplitude of the piezoelectric bender is depended on the piezoelectric coefficients and quality factor of the piezoelectric bender \cite{wang98}
	%	Improving vibration amplitude of mechanical oscillator is another way to improve modulation efficiency.
	%	Based on knowledge of engineering mechanics and material mechanics,
	%	Extra gap induced by nonzero dihedral angle between surface of modulation chip and FCs can be reduced by using goniometer stages for adjustment during installation step.
	According to the above discussion, the following methods can enhance the modulation efficiency.
	The first one is polishing of the modulation chip and the FCs, and second one is improving the vibration amplitude of the piezoelectric bender.
	The polishing of the modulation chip and the FCs enables the perfect contact of the two objects.
	The electron beam \cite{uno07} and micron-sized diamonds \cite{kubota20} could be used to reduce the surface roughness of the two objects to several nanometers.
	Under the same excitation voltage, the product of the vibration amplitude and the resonance frequency of the piezoelectric bender can be an approximate invariant, depended on piezoelectric coefficients and quality factor \cite{wang98}.
	PZT8 is a low damping piezoelectric material with high piezoelectric coefficients $d_{13}$ of 240 pC/N.
	%	With the metal bonding process, the quality factor of the PZT8 piezoelectric bender can reach 1189.
	As a reference, a piezoelectric bender is fabricated by cuboid cut PZT8 fastened to silicon substrate via Au-In metal bonding and realized resonance frequency over 8 kHz with vibration amplitude about 15 $\mu$m \cite{du19}.
	Thus a piezoelectric bender with the resonance frequency over 3 kHz and the vibration amplitude over 40 $\mu$m can be achievable by using PZT8.
	With the above improvements, the modulation efficiency about 56.8$\%$ can be achieved according the simulation as shown in Fig.~\ref{fig:S6}(d).
	%According to the requirement of the $T_{2}$,
	
	%	Because the designed vibration period of piezoelectric bender should be on the same order with diamond sample's $T_{2}$, the way to increase vibration amplitude is to find piezoelectric bender with higher piezoelectric coefficients and quality factor.

	%	The modulation efficiency achieved by this piezoelectric bender was 68.7 $\%$.

	\subsection{Improvement of the coherence time ($T_{\rm coh}$)}
	
	For the NV-magnetometry with FCM, the coherence time $T_{2}$ is regarded as $T_{\rm coh}$.
	Due to the low applied magnetic field in this work, no $^{13}$C revival was obtained as shown in Fig. 1(a) of main text.
	Under the applied magnetic field of low strength, the natural abundance of $^{13}$C decrease $T_{2}$.
	%	The coherent modulation of NV spin-echo signal due to Larmor precession of $^{13}$C nuclear bath spins should be avoided.  this coherent modulation om NV spin-echo signal
	With $^{12}$C enrich technique, the effect of $^{13}$C can be suppressed \cite{Bauch2020}.
	
	According to the detailed work on diamond samples \cite{Barry2020, Bauch2018}, the expected $T_{2}$ of $\sim$700 $\mu$s is achievable with the nitrogen concentration of about 0.05 ppm.
	The nitrogen-to-NV conversion efficiency of the diamond used in this work is about 37.5\% (Element Six, DNV B1).
	The NV concentration of 0.019 ppm is reachable for the nitrogen concentration of about 0.05 ppm in the future.
	%	$T_{\rm coh}$ will be replaced by $T_{2}$ in discussion about FCM method in the rest of the text, for convenience.
	%	Diamond sample with long $T_{2}$ is necessary for realizing high sensitivity.
	%	Based on recent research on NV coherence time, the fitted nitrogen-independent contribution to $T_{2}$ is 694 $\pm$ 82 $\mu$s \cite{Bauch2020}. This limitation on $T_{2}$ can be realized in diamond sample with low nitrogen concentration ($\approx$ 0.05 ppm).
	
	%	Since the magnetic field sensing in this work was achieved under applied magnetic field with low strength, $^{13}$C abundance is a non-negligible point that affects $T_{2}$.
	%	In natural $^{13}$C abundance sample, Larmor precession of $^{13}$C nuclear bath spins will cause coherent modulation of NV spin-echo signal and $T_{2}$ should be acquired by fitting envelope of $^{13}$C revival \cite{Bauch2020}.

	\subsection{Improvement of the average number of photons detected per measurement ($N$)}

	The number $N$ is depended on the photoluminescence rate $R$ of single NV center and the the number of the NV centers being excited.
	$R$ can be estimated as \cite{Chapman2011}
	\begin{equation}
		R=R_{\infty}(k) \frac{I}{I+I_{s}(\sigma,k)} ,\label{func5}
	\end{equation}
	%	_{0}e^{-\beta L}   ;; which depends on fluorescence collection efficiency $\phi$
	where $R_{\infty}$ is maximum detected photoluminescence rate of single NV center. $I$ is the power density of the laser. $I_{s}$ is saturation intensity of the laser.
	Because of the NV centers' absorption to laser, $I=I_{0}e^{-\beta L}$ decreases with the path length $L$ of excitation laser, where $I_{0}$ is the incident power density of the laser. $\beta$ is the absorption coefficient for the laser.
	The absorption cross-section $\sigma_{\rm NV}$ of NV center is about 1 $\times$ 10$^{-16}$ cm$^{2}$ \cite{Chapman2011}.
	Thus, $N$ can be estimated by the integral of $R$ for all NV centers being excited in the laser, and we have
	
	\begin{equation}
		N \approx \int_{0}^{L_{ \rm max}} R t_{r} n_{_{\rm NV}}\pi r^{2} dL ,\label{func6}
	\end{equation}
	where $L_{ \rm max}$ is the maximum path length of the excitation laser in the diamond. $t_{r}$ is the readout time per measurement. $n_{\rm NV}$ is the concentration of the NV centers in diamond. $r$ is the radius of the laser spot.
	
	In this work, the laser power was 0.375 W. $t_{r}$ was 9 $\mu$s. $r$ was 20 $\mu$m. The area of the laser spot was about $1.3\times10^{3}$ $\mu$m$^{2}$. $n_{\rm NV}$ was about 0.3 ppm. The attenuation coefficient $\beta=\sigma_{\rm NV} n_{\rm NV}$ was 528 m$^{-1}$.
	$L_{ \rm max}$ was 1 mm for the single-pass configuration.
	According to the experiment, the fluorescence intensity is about 0.22 mW.
	The average wavelength of the fluorescence is 680 nm.
	To fit the experimental result, $R_{\infty}=44.8$ kHz is obtained with the assumption of $I_{s}=650$ MW/m$^{2}$ \cite{plakhotnik10}.
	Thus $N$ is calculated as 7.0 $\times$ 10$^{9}$.
	
	%intensity of laser decrease to 1\% of the incident intensity
	To improve $N$, we are going to use the total internal reflection \cite{Clevenson2015} to increase the number of the NV centers being excited.
	With $n_{\rm NV}$ of 0.019 ppm in the future, we have the attenuation coefficient $\beta=\sigma_{\rm NV} n_{\rm NV}$ of 33.44 m$^{-1}$.
	$L_{\rm max}$ will be $\sim$32 mm, while the whole diamond with the size of 0.04 mm $\times$ 1 mm $\times$ 1 mm is excited.
	According to the simulation result, the collection efficiency can be enhanced by 2.2 folds with the coating technique \cite{yu20} and a better fabrication technique of diamond and CPC.
	Thus $R_{\infty}$ of 98.6 kHz is available with the improvements.
	It should be noticed that a long-pass dielectric film is required to prevent the intensity loss of the laser.
	The dielectric film should be able to reflect the laser, and coated on the diamond surface contacted with CPC.
	To avoid effects on other parameters, we do not change the laser power and the diameter of the laser spot in the calculation.
	With $I_{s}=650$ MW/m$^{2}$ \cite{plakhotnik10} and $R_{\infty}=98.6$ kHz, the $N$ can be improved from 7.0 $\times$ 10$^{9}$ to 2.6 $\times$ 10$^{10}$.
	%It is worth mentioning that the increase of the laser power could lead to the many folds enhancement of $N$.
	%The optimization
	
	\subsection{Improvement of the measurement contrast ($C$)}

	For the Ramsey-type NV-magnetometry, $C$ was about 1.2$\times$10$^{-2}$.
	For the NV-magnetometry with FCM, $C$ was about 4.5$\times$10$^{-3}$.
	The decrease mainly comes from the magnetic field inhomogeneity in diamond.
	We describe a model for the impact of the magnetic field inhomogeneity.
	%	The total magnetic field acting on the ensemble of NV centers can be represented as
	%	The magnetic field acting on the NV center can be represented as +B_{\Delta})
	The signal of the magnetometry based on the ensemble of NV centers can be represented as
	
	\begin{equation}
		S= \int dG' \int dB_{r} P_0(G') P_1(B_{r}) S'(B)  ,
	\end{equation}
	where $B=G' B_{a}+B_{r}$ is the magnetic field acting on the NV centers, and $S'(B)$ is the readout signal of the NV centers under $B$. $G'$ is the magnification of the FCs for $B_{a}$. $B_{a}$ is the applied magnetic field to the probe. $B_{r}$ is a static magnetic field, which could come from the remanence of the FCs. $P_0(G')$ and $P_1(B_{r})$ are the distributions of the parameters.
	
	According to previous discussions in the solid-state spins \cite{Bauch2020,Rong2015,Boss2017}, we suppose that $P_0(G')$ and $P_1(B_{r})$ are satisfied with the widely used normal distribution. Thus the two distributions are characterized as 
	%	$x_{0}$ and $\sigma$ are average and scale parameter of normal distribution.
	\begin{equation}\label{key}
		\begin{aligned}
			\begin{cases}
				P_0(G') &= \frac{1}{\sqrt{2\pi} kG}exp\left[-\frac{(G'-G)^{2}}{2(kG)^{2}}\right] \\
				%			P_1(B_{\Delta}) &=\frac{1}{\sqrt{2\pi}l B_{\Delta}}exp\left[-\frac{(B_{\Delta}-B_{\Delta 0})^{2}}{2(l B_{\Delta})^{2}}\right] \\
				P_1(B_{r}) &= \frac{1}{\sqrt{2\pi} mB_{r0}}exp\left[-\frac{(B_{r}-B_{r0})^{2}}{2(mB_{r0})^{2}}\right] \\
			\end{cases}
		\end{aligned}
	\end{equation}
	where $k$ and $m$ are relative scale parameters of the corresponding distributions.
	$G$ and $B_{r0}$ are mean values of the corresponding distributions. 
	
	%	$kG$ is assumed to adjust distribution $\sigma$ of $G$.
	%	$G$ is averaged magnification of the FCs here, as defined in section \ref{sec-G}.
	%	$k$ is a manual parameter related to the inhomogeneity of $G$.
	%	$mB_{r0}$ is used to adjust distribution $\sigma$ of $B_{r}$. 
	%	$B_{r0}$ is averaged remanence of the FCs.
	%	$m$ is a manual parameter related to the inhomogeneity of $B_{r}$.
	%	The distribution of $B_{\Delta}$ is assumed to satisfy with uniform distribution of $p(x)={1}/{(b-a)}$.
	%	$a$ and $b$ are lower bound and upper bound of the uniform distribution.
	
	\begin{figure*}[!htp]
		\includegraphics[width=0.6\textwidth]{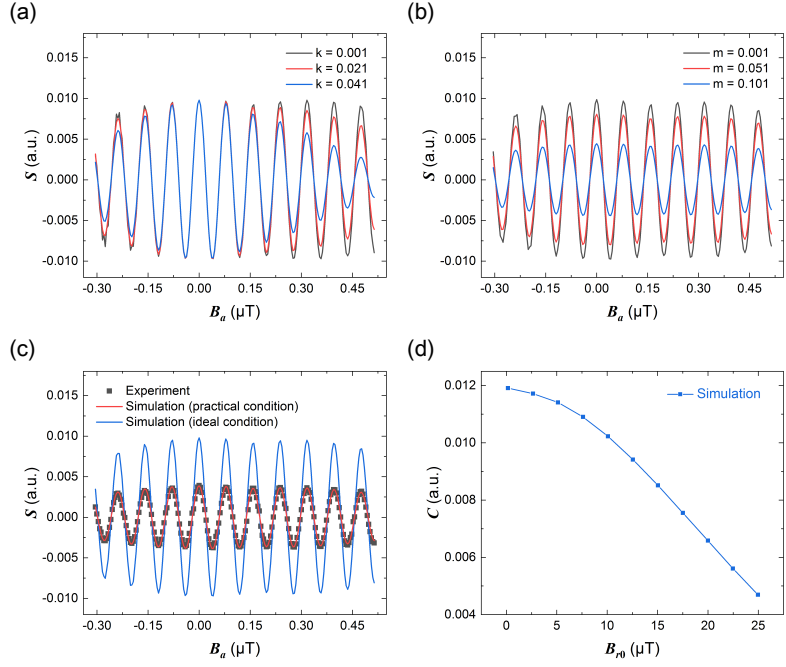}% Here is how to import EPS art
		\caption{\label{fig:S7} The simulation of $C$ under different conditions. (a) The signal as the function of $B_{a}$ under different $k$. Factor $m$ was set to be 0.01 in the simulation. (b) The signal as the function of $B_{a}$ under different $m$. Factor $k$ was set to be 0.01 in the simulation. (c) The comparison between the experiment and the simulation. In the simulation with the practical condition, $k$ and $m$ were set to be 0.015 and 0.109, respectively. In the simulation with the ideal condition, $k$ and $m$ were both set to be 0.01. (d) The simulation result of $C$ as the function of the averaged remanence of the FCs $B_{r0}$. $k$ and $m$ were set to be 0.015 and 0.109 as in the simulation of the practical condition.}
	\end{figure*}
	
	%	magnetic field response simulated
	The simulation is performed by changing applied magnetic field $B_{a}$.
	%	$k$ and $m$ are the main parameters need to be adjusted for acquiring fitness between experiment and simulation.
	Fig.~\ref{fig:S7}(a) and Fig.~\ref{fig:S7}(b) show the signal as the function of $B_{a}$ with different $k$ and $m$, respectively.
	The contrast of the signal decreases with factor $k$ increasing at the large $B_{a}$.
	The holistic decrease of the contrast can be observed with the increase of $m$.
	It can be summarized as that the stronger magnetic field inhomogeneity leads to the larger decrease of $C$.
	
	To match with the experimental data, we adjust $k$ and $m$.
	Fig.~\ref{fig:S7}(c) shows the comparison between the experiment and the simulation, which is perfectly matching.
	It should be noticed that the spectra is not symmetry about $B_{a}=0$ due to $B_{r0}$.
	All the parameters used in the simulation are listed in Table~\ref{tab:table2}.
	
	%	It is not difficult to comprehend this result.
	%	The contrast decrease is induced by inhomogeneity of total magnetic field acting on NV ensembles.
	%	As mentioned before, the total magnetic field acting on NV ensembles can be written as $B(t)=G(B_{a}+B_{\Delta})+B_{r}$.
	
	%	Notice the contrast decrease contributed by inhomogeneity of $G$ depends on $B_{a}$.
	%	Only if $B_{a}$ is strong enough, significant contrast decrease can be observed.
	
	%	The contrast decrease at high applied magnetic field $B_{a}$ depends on factor $k$ which decides the inhomogeneity of $G$.
	%	However, FC's remanence $B_{r}$ doesn't change with $B_{a}$ and contrast decrease caused by inhomogeneity of $B_{r}$ has a holistic effect on the spectrum.
	%	The holistic contrast decrease depends on factor $m$ which decides the inhomogeneity of $B_{r}$.

	%	The maximum peak-to-peak value in signal to magnetic field response was mainly affected by factor $m$ as shown in Fig.~\ref{fig:S7}(a) and Fig.~\ref{fig:S7}(b).
	%	So the decrease of measurement contrast is mainly induced by inhomogeneity of FC's remanence, for the magnetic field sensing realized under applied magnetic field with low strength in the work. This inhomogeneity is the inherent attribute of 1J85 alloy and it is difficult to be eliminated.
	%However, because the inhomogeneity comes from FC's remanence, a more appropriate way to suppress its impact is eliminating FC's remanence directly.
	According to the above simulations, the reduction of FC's remanence is helpful in the improvement of $C$.
	Fig.~\ref{fig:S7}(d) shows the simulation result of $C$ as the function $B_{r0}$.
	To increase $C$ to about 1.2$\times$10$^{-2}$, we have to suppress $B_{r0}$ to less than 2 $\mu$T from Fig.~\ref{fig:S7}(d).
	The elimination of FC's remanence can be realized by applying an ac magnetic field to the FC \cite{gerdroodbari18}.
	%	The AC magnetic field can disorganize the configuration of magnetic domain in FC and eliminate FC's remanence. Since the system is aimed at measuring weak magnetic field, the elimination of FC's remanence can be effective for a long time. With elimination of FC's remanence, the measurement contrast can be recovered to $\sim$0.01.

	\begin{table*}[!htp]
		\caption{\label{tab:table2}Parameters used in simulation for Fig.~\ref{fig:S7}(c)}
		\begin{ruledtabular}
			\begin{tabular}{ccccc}
				Parameter&Value&Parameter&Value\\ \hline
				$k$&0.015 (practical) / 0.01 (ideal)&$G$&85.1\\
				$m$&0.109 (practical) / 0.01 (ideal)&$B_{r0}$&25.0 $\mu$T\\
				$C$&1.2 $\times$ 10$^{-2}$&$T_{2}$&102 $\mu$s\\
				$p$&1.24&$\tau$&92.7 $\mu$s\\
			\end{tabular}
		\end{ruledtabular}
	\end{table*}
	
	\subsection{The potentially achievable sensitivity}
	%	 and its shot noise limitation is $\sim$3 fT/Hz$^{1/2}$, with the improvements on several parameters mentioned before
	According to the above discussions, the shot-noise-limited sensitivity of the NV-magnetometry with FCM can reach $\sim$3 fT/Hz$^{1/2}$, evaluated by the equation (\ref{eq-pulse}).
	For an achievable $n_f$ of 3, the sensitivity of femtotesla level is promising.
	All the parameters used in the evaluation are listed in Table~\ref{tab:table3}.
	
	From Table~\ref{tab:table3}, we can see the essential enhancements coming from the magnification $G$ of the FCs, the modulation efficiency $E_{_F}$, and the noise ratio $n_f$. The improvements of $G$ and $E_{_F}$ require the micromachining technologies. The improvement of $n_f$ requires further efforts on the readout system optimization.
	Although a sensitivity of femtotesla level is proposed, more efforts on the excitation and the quantum sensing protocols could lead to a more promising result.
	This $T_2$-limited dc quantum magnetometry is worth further exploration.
	
	\begin{table*}[!htp]
		\caption{\label{tab:table3}Parameters used in sensitivity calculation.}
		\begin{ruledtabular}
			\begin{tabular}{ccccc}
				Parameter&Present value&Improved value&Enhancement for sensitivity&Reference\\ \hline
				$G$&85.1&527&6.2&Simulation\\
				$E_{_F}$&0.096&0.568&5.9&\cite{uno07}, \cite{kubota20}, \cite{du19}, Simulation\\
				$T_{2}$&102 $\mu$s&$\sim$ 694 $\mu$s&$\sim$ 1.6&\cite{Bauch2020}\\
				$N$&7.0 $\times$ 10$^{9}$&2.6 $\times$ 10$^{10}$&1.9&\cite{Clevenson2015}, \cite{yu20}\\
				$C$&4.5 $\times$ 10$^{-3}$&$\sim$ 1.2 $\times$ 10$^{-2}$&$\sim$ 2.7&\cite{gerdroodbari18}, Simulation\\
				$\tau$&92.7 $\mu$s&333.6 $\mu$s&2.5&Experiment\\
				$t_{m}$&140 $\mu$s&140 $\mu$s&$/$&Experiment\\
				$n_{f}$&19.2&19.2&$/$&Experiment\\
				$p$&1.24&1.24&$/$&Experiment\\
				$\gamma_{e}$&2$\pi$ $\times$ 28 GHz/T&2$\pi$ $\times$ 28 GHz/T&$/$&$/$\\
				$\alpha$&0.5774&0.5774&$/$&$/$\\ \hline
				$\eta$ (Evaluation)&39 pT/Hz$^{1/2}$&$\sim$ 50 fT/Hz$^{1/2}$&$\sim$ 750.6&Calculation\\
				$\eta$ (Shot-noise-limited)&2 pT/Hz$^{1/2}$&$\sim$ 3 fT/Hz$^{1/2}$&$\sim$ 750.6&Calculation\\
			\end{tabular}
		\end{ruledtabular}
	\end{table*}
	
%	\clearpage
	%\nocite{*}%
%	\bibliography{ref, Supplementarymaterial}% Produces the bibliography via BibTeX.

\begin{thebibliography}{10}
		
		\bibitem{Boto2018}
		E.~Boto, {\it et~al.\/}, {Moving magnetoencephalography towards real-world
			applications with a wearable system}. {\it Nature\/} {\bf 555}, 657 (2018).
		
		\bibitem{Cohen1967}
		D.~Cohen, {Magnetic Fields around the Torso: Production by Electrical Activity
			of the Human Heart}. {\it Science\/} {\bf 156}, 652 (1967).
		
		\bibitem{Lenz2006}
		J.~Lenz, S.~Edelstein, {Magnetic sensors and their applications}. {\it IEEE
			Sensors Journal\/} {\bf 6}, 631 (2006).
		
		\bibitem{Glenn2017}
		D.~R. Glenn, {\it et~al.\/}, {Micrometer-scale magnetic imaging of geological
			samples using a quantum diamond microscope}. {\it Geochemistry, Geophysics,
			Geosystems\/} {\bf 18}, 3254 (2017).
		
		\bibitem{Jiao2021}
		M.~Jiao, M.~Guo, X.~Rong, Y.-F. Cai, J.~Du, {Experimental Constraint on an
			Exotic Parity-Odd Spin- and Velocity-Dependent Interaction with a Single
			Electron Spin Quantum Sensor}. {\it Physical Review Letters\/} {\bf 127},
		010501 (2021).
		
		\bibitem{Degen2017}
		C.~L. Degen, F.~Reinhard, P.~Cappellaro, {Quantum sensing}. {\it Reviews of
			Modern Physics\/} {\bf 89}, 1 (2017).
		
		\bibitem{Budker2007}
		D.~Budker, M.~Romalis, {Optical magnetometry}. {\it Nature Physics\/} {\bf 3},
		227 (2007).
		
		\bibitem{Dolabdjian2017}
		C.~Dolabdjian, D.~Menard, A.~Grosz, M.~J. Haji-Sheikh, S.~C. Mukhopadhyay, {\it
			{High Sensitivity Magnetometers}\/}, vol.~19 of {\it Smart Sensors,
			Measurement and Instrumentation\/} (Springer International Publishing, Cham,
		2017).
		
		\bibitem{Taylor2008}
		J.~M. Taylor, {\it et~al.\/}, {High-sensitivity diamond magnetometer with
			nanoscale resolution}. {\it Nature Physics\/} {\bf 4}, 810 (2008).
		
		\bibitem{Jensen2014}
		K.~Jensen, {\it et~al.\/}, {Cavity-Enhanced Room-Temperature Magnetometry Using
			Absorption by Nitrogen-Vacancy Centers in Diamond}. {\it Physical Review
			Letters\/} {\bf 112}, 160802 (2014).
		
		\bibitem{Chatzidrosos2017}
		G.~Chatzidrosos, {\it et~al.\/}, {Miniature Cavity-Enhanced Diamond
			Magnetometer}. {\it Physical Review Applied\/} {\bf 8}, 044019 (2017).
		
		\bibitem{Clevenson2015}
		H.~Clevenson, {\it et~al.\/}, {Broadband magnetometry and temperature sensing
			with a light-trapping diamond waveguide}. {\it Nature Physics\/} {\bf 11},
		393 (2015).
		
		\bibitem{Xie2021}
		Y.~Xie, {\it et~al.\/}, {A hybrid magnetometer towards femtotesla sensitivity
			under ambient conditions}. {\it Science Bulletin\/} {\bf 66}, 127 (2021).
		
		\bibitem{Fescenko2020}
		I.~Fescenko, {\it et~al.\/}, {Diamond magnetometer enhanced by ferrite flux
			concentrators}. {\it Physical Review Research\/} {\bf 2}, 023394 (2020).
		
		\bibitem{Zhang2021c}
		C.~Zhang, {\it et~al.\/}, {Diamond Magnetometry and Gradiometry Towards
			Subpicotesla dc Field Measurement}. {\it Physical Review Applied\/} {\bf 15},
		064075 (2021).
		
		\bibitem{Barry2017}
		F.~Barry, {\it et~al.\/}, {Optical magnetic detection of single-neuron action
			potentials using quantum defects in diamond}. {\it Proceedings of the
			National Academy of Sciences\/} {\bf 114}, E6730 (2017).
		
		\bibitem{Barry2020}
		J.~F. Barry, {\it et~al.\/}, {Sensitivity optimization for NV-diamond
			magnetometry}. {\it Reviews of Modern Physics\/} {\bf 92}, 015004 (2020).
		
		\bibitem{Bauch2018}
		E.~Bauch, {\it et~al.\/}, {Ultralong Dephasing Times in Solid-State Spin
			Ensembles via Quantum Control}. {\it Physical Review X\/} {\bf 8}, 031025
		(2018).
		
		\bibitem{Wolf2015}
		T.~Wolf, {\it et~al.\/}, {Subpicotesla Diamond Magnetometry}. {\it Physical
			Review X\/} {\bf 5}, 041001 (2015).
		
		\bibitem{Liu2019a}
		Y.-X. Liu, A.~Ajoy, P.~Cappellaro, {Nanoscale Vector dc Magnetometry via
			Ancilla-Assisted Frequency Up-Conversion}. {\it Physical Review Letters\/}
		{\bf 122}, 100501 (2019).
		
		\bibitem{Wood2018a}
		A.~A. Wood, {\it et~al.\/}, {T2-limited sensing of static magnetic fields via
			fast rotation of quantum spins}. {\it Physical Review B\/} {\bf 98}, 174114
		(2018).
		
		\bibitem{Hahn1950}
		E.~L. Hahn, Spin echoes. {\it Phys. Rev.\/} {\bf 80}, 580 (1950).
		
		\bibitem{Tian2013}
		W.~Tian, J.~Hu, M.~Pan, D.~Chen, J.~Zhao, {Flux concentration and modulation
			based magnetoresistive sensor with integrated planar compensation coils}.
		{\it Review of Scientific Instruments\/} {\bf 84}, 035004 (2013).
		
		\bibitem{Edelstein2006}
		A.~S. Edelstein, {\it et~al.\/}, {Progress toward a thousandfold reduction in
			1/f noise in magnetic sensors using an ac microelectromechanical system flux
			concentrator (invited)}. {\it Journal of Applied Physics\/} {\bf 99}, 08B317
		(2006).
		
		\bibitem{Guedes2008}
		A.~Guedes, {\it et~al.\/}, {Hybrid magnetoresistive/microelectromechanical
			devices for static field modulation and sensor 1/f noise cancellation}. {\it
			Journal of Applied Physics\/} {\bf 103}, 1 (2008).
		
		\bibitem{Bayat2014}
		K.~Bayat, J.~Choy, M.~{Farrokh Baroughi}, S.~Meesala, M.~Loncar, {Efficient,
			Uniform, and Large Area Microwave Magnetic Coupling to NV Centers in Diamond
			Using Double Split-Ring Resonators}. {\it Nano Letters\/} {\bf 14}, 1208
		(2014).
		
		\bibitem{Qin2020}
		X.~Qin, {\it et~al.\/}, An fpga-based hardware platform for the control of
		spin-based quantum systems. {\it IEEE Transactions on Instrumentation and
			Measurement\/} {\bf 69}, 1127 (2020).
		
		\bibitem{Zheng2019}
		H.~Zheng, {\it et~al.\/}, {Zero-Field Magnetometry Based on Nitrogen-Vacancy
			Ensembles in Diamond}. {\it Physical Review Applied\/} {\bf 11}, 064068
		(2019).
		
		\bibitem{Zheng2020}
		D.~Zheng, {\it et~al.\/}, {A hand-held magnetometer based on an ensemble of
			nitrogen-vacancy centers in diamond}. {\it Journal of Physics D: Applied
			Physics\/} {\bf 53}, 155004 (2020).
		
		\bibitem{Du2019}
		Q.~Du, {\it et~al.\/}, {High Efficiency Magnetic Flux Modulation Structure for
			Magnetoresistance Sensor}. {\it IEEE Electron Device Letters\/} {\bf 40},
		1824 (2019).
		
		\bibitem{Schloss2018}
		J.~M. Schloss, J.~F. Barry, M.~J. Turner, R.~L. Walsworth, {Simultaneous
			Broadband Vector Magnetometry Using Solid-State Spins}. {\it Physical Review
			Applied\/} {\bf 10}, 034044 (2018).
		
		\bibitem{Kim2019}
		D.~Kim, {\it et~al.\/}, {A CMOS-integrated quantum sensor based on
			nitrogen–vacancy centres}. {\it Nature Electronics\/} {\bf 2}, 284 (2019).
		
		\bibitem{pan12}
		M.~Pan, J.~Hu, W.~Tian, D.~Chen, J.~Zhao, Magnetic flux vertical motion
		modulation for 1/f noise reduction of magnetic tunnel junctions. {\it Sensors
			and Actuators A: Physical\/} {\bf 179}, 92 (2012).
		
		\bibitem{uno07}
		Y.~Uno, {\it et~al.\/}, A new polishing method of metal mold with large-area
		electron beam irradiation. {\it Journal of Materials processing technology\/}
		{\bf 187}, 77 (2007).
		
		\bibitem{kubota20}
		A.~Kubota, S.~Nagae, S.~Motoyama, High-precision mechanical polishing method
		for diamond substrate using micron-sized diamond abrasive grains. {\it
			Diamond and Related Materials\/} {\bf 101}, 107644 (2020).
		
		\bibitem{wang98}
		Q.-M. Wang, L.~E. Cross, Performance analysis of piezoelectric cantilever
		bending actuators. {\it Ferroelectrics\/} {\bf 215}, 187 (1998).
		
		\bibitem{du19}
		Q.~Du, {\it et~al.\/}, High efficiency magnetic flux modulation structure for
		magnetoresistance sensor. {\it IEEE Electron Device Letters\/} {\bf 40}, 1824
		(2019).
		
		\bibitem{Bauch2020}
		E.~Bauch, {\it et~al.\/}, {Decoherence of ensembles of nitrogen-vacancy centers
			in diamond}. {\it Physical Review B\/} {\bf 102}, 134210 (2020).
		
		\bibitem{Chapman2011}
		R.~Chapman, T.~Plakhotnik, {Quantitative luminescence microscopy on
			Nitrogen-Vacancy Centres in diamond: Saturation effects under pulsed
			excitation}. {\it Chemical Physics Letters\/} {\bf 507}, 190 (2011).
		
		\bibitem{plakhotnik10}
		T.~Plakhotnik, D.~Gruber, Luminescence of nitrogen-vacancy centers in
		nanodiamonds at temperatures between 300 and 700 k: perspectives on
		nanothermometry. {\it Physical Chemistry Chemical Physics\/} {\bf 12}, 9751
		(2010).
		
		\bibitem{yu20}
		H.~Yu, Y.~Xie, Y.~Zhu, X.~Rong, J.~Du, Enhanced sensitivity of the
		nitrogen-vacancy ensemble magnetometer via surface coating. {\it Applied
			Physics Letters\/} {\bf 117}, 204002 (2020).
		
		\bibitem{Rong2015}
		X.~Rong, {\it et~al.\/}, {Experimental fault-tolerant universal quantum gates
			with solid-state spins under ambient conditions}. {\it Nature
			Communications\/} {\bf 6}, 8748 (2015).
		
		\bibitem{Boss2017}
		J.~M. Boss, K.~S. Cujia, J.~Zopes, C.~L. Degen, {Quantum sensing with arbitrary
			frequency resolution}. {\it Science\/} {\bf 356}, 837 (2017).
		
		\bibitem{gerdroodbari18}
		Y.~Z. Gerdroodbari, M.~Davarpanah, S.~Farhangi, Remanent flux negative effects
		on transformer diagnostic test results and a novel approach for its
		elimination. {\it IEEE Transactions on Power Delivery\/} {\bf 33}, 2938
		(2018).
		
	\end{thebibliography}
	
\end{document}